\RequirePackage{fix-cm}
\documentclass[smallextended]{svjour3}       % onecolumn (second format)
\smartqed  % flush right qed marks, e.g. at end of proof
\usepackage{graphicx}

\usepackage[utf8]{inputenc}%Einbinden der Umlaute
\usepackage{siunitx}
\usepackage{nameref}

\usepackage{amsfonts}%Mathematische Zeichen
\usepackage{amsmath}%Matrizen, etc.

\usepackage{graphicx}
\usepackage{tikz}
\usetikzlibrary{shapes,decorations.pathmorphing, external}
\usepackage{pgfplots}
\usepgfplotslibrary{groupplots, statistics}
\pgfplotsset{compat=newest}
\usepackage{placeins}
\usepackage{mathtools}
\usepackage{pifont}

\usepackage{algorithm}
\usepackage{algorithmic}
\usepackage{float}
\newfloat{algorithm}{tb}{lop}

\usepackage{xcolor}

\usepackage{relsize}

\usepackage{xspace}

\usepackage[hidelinks]{hyperref}
 
\newcommand{\abs}[1]{\ensuremath{\left\vert#1\right\vert}}
\newcommand{\norm}[1]{\left\lVert#1\right\rVert}

\newcommand{\sign}[1]{\ensuremath{\mathrm{sign}\left(#1\right)}}
\newcommand{\diag}[1]{\ensuremath{\mathrm{diag}\left[#1\right]}}

\newcommand{\dt}{\ensuremath{\ \mathrm{d}t}}

\newcommand{\V}{\ensuremath{\mathcal{V}}}
\newcommand{\st}{\ensuremath{\mathrm{s.t.\ }}}

\newcommand{\R}{\ensuremath{\mathbb{R}}}
\renewcommand{\vec}[1]{\boldsymbol{#1}}
\newcommand{\TT}{\ensuremath{\mathsf{T}}}
\newcommand{\N}{\ensuremath{\mathcal{N}}}

\newcommand{\sending}{\ensuremath{\leftarrow}}
\newcommand{\receiving}{\ensuremath{\rightarrow}}
\newcommand{\shortminus}{\text{-}}

\newcommand{\ADMM}{\ensuremath{\mathrm{ADMM}}\xspace}
\newcommand{\GRAMPC}{\textrm{GRAMPC}\xspace}
\newcommand{\GRAMPCD}{\mbox{\textrm{GRAMPC-D}}\xspace}
\newcommand{\MPC}{\ensuremath{\mathrm{MPC}}\xspace}
\newcommand{\DMPC}{\ensuremath{\mathrm{DMPC}}\xspace}
\newcommand{\OCP}{\ensuremath{\mathrm{OCP}}\xspace}

\newcommand{\ALADIN}{\ensuremath{\mathrm{ALADIN}}\xspace}

\newcommand{\widthAgent}{1.5cm}
\newcommand{\linewidthfigure}{0.75}
\newcommand{\widthPlot}{10cm}
\newcommand{\heightPlot}{4cm}

\newcommand{\Cpp}{C\nolinebreak\hspace{-.05em}\raisebox{.4ex}{\tiny\bf +}\nolinebreak\hspace{-.10em}\raisebox{.4ex}{\tiny\bf +}\xspace}

\newcommand{\MATLAB}{\textsc{Matlab}\xspace}

\RequirePackage{color}
\definecolor{myred}{rgb}{1,0,0}
\definecolor{myblue}{rgb}{0,0,1}

\begin{document}

\title{A Modular Framework for Distributed Model Predictive Control of Nonlinear Continuous-Time Systems (\GRAMPCD)}
\titlerunning{A Modular Framework for \DMPC}

\author{Daniel Burk \and
        Andreas Völz \and
        Knut Graichen%
}

\institute{
	Friedrich-Alexander-Universität Erlangen-Nürnberg \\
	Lehrstuhl für Regelungstechnik \at
	Cauerstraße 7, 91058 Erlangen \\
	\email{daniel.burk@fau.de}     
}

\date{Received: date / Accepted: date}

\maketitle

\begin{abstract}
  The modular open-source framework \GRAMPCD for model predictive control
  of distributed systems is presented in this paper. The modular
  concept allows to solve optimal control problems (OCP) in
  a centralized and distributed fashion using the same problem
  description. It is tailored to computational efficiency with the focus
  on embedded hardware. The distributed solution is based on the
  Alternating Direction Method of Multipliers (\ADMM) and uses the
  concept of neighbor approximation to enhance convergence speed. The
  presented framework can be
  accessed through \Cpp\ and Python and also supports plug-and-play and data
  exchange between agents over a network. 
\end{abstract}

\section{Introduction}

Model predictive control (MPC) is a modern control concept that
attained increasing attention during the last decades
\cite{bib:mayne2000,bib:allgower2012} as it is capable to handle
nonlinear systems while considering constraints on both states and
controls. It is based on solving an optimal control problem (OCP) on a
finite horizon and applying the first part of the control trajectory
%$\Delta_t$ seconds of the trajectory on 
to the actual plant, corresponding to the sampling time $\Delta_t$ of
the controller. At the next sampling instant, the
horizon is shifted and the OCP is solved again. This iterative scheme
is executed repetitively to stabilize the plant on an infinite
horizon.  

A main difficulty is the computational complexity of solving the OCP
in real-time, which in turn requires an efficient implementation of
suitable MPC algorithms. In the recent past, several toolboxes were
published that provide adequate software frameworks such as ACADO 
\cite{bib:houska2011} and ACADOS \cite{bib:verschueren2019}, VIATOC
\cite{bib:kalmari2015} or \GRAMPC\
\cite{bib:kapernick2014,bib:englert2019}. In case of distributed
systems with a high number of controls and states, the
classic centralized approach is not capable of solving the overall OCP
in real-time anymore. Hence, algorithms for distributed model
predictive control (DMPC) \cite{bib:camponogara2002,bib:maestre2014}
have been in the focus over the last years. Their basic idea is to
decouple the centralized OCP 
and to split it into multiple local OCPs that can be solved in
parallel. The expectation is to compensate the higher computational
complexity due to the decoupled formulation as well as the increased
communication effort by the parallel structure. There are multiple
approaches to distributed algorithms for optimal control problems,
such as sensitivity-based algorithms \cite{bib:scheu2011}, the
augmented  Lagrangian  based  alternating  direction  inexact  Newton
method (\ALADIN) \cite{bib:houska2016,bib:houska2018} or the
alternating direction method of multipliers (\ADMM)
\cite{bib:boyd2011} that is also used in the presented framework.

The difficulty of an efficient implementation is drastically higher in
case of DMPC than for classic MPC algorithms, as a potentially high
number of subsystems, so-called \textit{agents}, have to be managed.
Several toolboxes for DMPC have been published as well. Linear 
discrete-time systems are considered in 
the DMPC-Toolbox \cite{bib:gafvert2014} that is implemented in
\MATLAB. The PnPMPC-TOOLBOX \cite{bib:pnpmpctoolbox2013} focuses on
the plug-and-play functionality and provides an implementation in
\MATLAB that considers continuous-time and discrete-time linear
systems. Several algorithms are implemented in the Python-Toolbox
DISROPT \cite{bib:farina2019} regarding distributed optimization
problems. ALADIN-$\alpha$ \cite{bib:engelmann2020} is the most recent
published toolbox that provides a \MATLAB implementation of the
ALADIN algorithm. However, there
is a lack of a DMPC framework that provides an implementation tailored
to embedded hardware with the focus on real-time capable distributed
model predictive control. Many real-world problems such as
smart grids or cooperative robot applications are only equipped with
weak hardware on the subsystem level that is not able to handle complex
computation tasks in an appropriate time. Hence, for realizing
distributed controllers on actual plants, an implementation optimized
on execution time is required to enable real-time control. Furthermore,
providing the possibility of communication between agents over a network 
is essential for a DMPC framework designed to control actual plants. 
The restriction to neighbor-to-neighbor-communication decouples the agents communication 
effort from the overall system size by bounding it to the cardinality
of its neighborhood. The focus on real-world plants requires the system class
to cover nonlinear dynamics including couplings between the agents in both
dynamics and constraints.

The presented framework, in the following \GRAMPCD,
provides an open-source \Cpp implementation that is capable of
solving optimal control problems in a distributed manner with a
per-agent computation-time in the millisecond range. The 
underlying minimization problems are solved with the \MPC toolbox
\GRAMPC that is suitable for embedded hardware implementations. However, other toolboxes for solving the local minimization problem can be used as well. To
enable actual distributed 
optimization, a socket-based TCP communication is provided to allow
agents to exchange data over a network. Furthermore, a
Python interface is provided in addition to the \Cpp-interface using
the software module Pybind11 \cite{bib:pybind11}. The Python interface
combines both the functionality of Python and the performance of \Cpp
as it only wraps the \Cpp interface while the actual code executing
the \DMPC algorithm is still running in \Cpp. Furthermore, it allows
for fast and efficient prototyping when
developing a controller for a distributed system as both a
centralized as well as a distributed controller can be derived based
on the same problem description. The convergence behavior of distributed
controllers can be improved by optionally using the concept of 
neighbor approximation. Thereby, the generated problem description
of each agent is adapted to additionally approximate parts of its neighbors
\OCP and by this to improve the solution of its local \OCP in each iteration,
leading to an enhanced convergence behavior of the overall algorithm.
The modular structure of \GRAMPCD enables modifying
the overall system in the sense of plug-and-play by including or removing agents
or couplings at run-time. Supporting plug-and-play features is a
core functionality for a DMPC framework with focus on embedded systems,
as the assumption of a static system
description does not hold for a large number of
real-world plants.

The paper is structured as follows. 
%The mathematical notation in the paper is shortly introduced in Section \ref{sec:notation}. 
Section
\ref{sec:problem_description} outlines the considered class of coupled
systems and OCP formulation. The DMPC framework \GRAMPCD is introduced in 
Section \ref{sec:DMPC} including the \ADMM algorithm as the
method of choice for the algorithm. In addition, the concept of
neighbor approximation is explained and the implemented algorithm for
the crucial task of penalty parameter adaption is presented. The
modular structure of the 
framework is presented in Section \ref{sec:framework}. Finally,
simulation examples in Section \ref{sec:simulation_example} show the
effectiveness and modularity of the DMPC framework, before conclusions
are drawn in Section~\ref{sec:conclusions}.

%\section{Notation}
%\label{sec:notation}

Throughout the paper, each
%Each 
vector $\vec x\in\mathbb{R}^n$ is written in bold style. Standard 
$p$-norms $\norm{\vec x}_p=\left( \sum_{i=1}^n
 \abs{x_i}^p \right)^{\frac{1}{p}}$ will be used as well as the weighted squared
norm defined by $\norm{\vec x}_{\vec P}^2 = \vec x^\TT \vec P \vec x$
with a positive (semi-)definite square matrix $\vec P$. 
The stacking of individual vectors $\vec x_i,\ i\in\V$ from a set $\V$
is denoted by $\vec x = \begin{bmatrix} \vec x_i \end{bmatrix}_{i\in\V}$.
As far as time trajectories are concerned, the explicit dependency on
time $t$ may be omitted to ease readability. The derivative with
respect to time is written using the dot notation $\vec{\dot x}(t) =
\frac{\operatorname{d}}{\operatorname{d}\!t}\vec x(t)$. 

\section{Problem Description}
\label{sec:problem_description}

The presented DMPC framework considers multi-agent systems that can be
described by a graph $\mathcal{G}=(\V, \mathcal{E})$ with the sets of edges
$\mathcal{E}$ and vertices $\mathcal{V}$. Each vertex represents an
agent, while each edge between two vertices stands for a coupling
between the corresponding agents. The couplings may be both uni- and
bi-directional. 

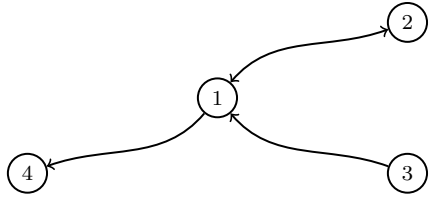
\begin{figure}
	\centering
	\begin{tikzpicture}
		% agents
		\node (A1) at (0, 0) [line width = \linewidthfigure, circle, draw] {$1$};
		\node (A2) at (2.5, 1) [line width = \linewidthfigure, circle, draw] {$2$};
		\node (A3) at (2.5, -1) [line width = \linewidthfigure, circle, draw] {$3$};
		\node (A4) at (-2.5, -1) [line width = \linewidthfigure, circle, draw] {$4$};
		% couplings
		\draw[line width = \linewidthfigure, <->] (A1) to[out = 50, in = 200] (A2);
		\draw[line width = \linewidthfigure, <-] (A1) to[out = -50, in = 160] (A3);
		\draw[line width = \linewidthfigure, ->] (A1) to[out = 180+50, in = 20] (A4);
	\end{tikzpicture}
	\caption{The neighborhood $\N_1=\{2, 3, 4\}$ of agent $1$ is composed of sending neighbors $\N_1^\sending=\{2, 3\}$ and receiving neighbors $\N_1^\receiving=\{2, 4\}$.  }
	\label{img:neighborhood}
\end{figure}

The considered optimal control problem for the coupled nonlinear
system is given by 
\begin{subequations}
	\begin{align}
		\min_{\vec u_i, i\in\V} \hspace{0.2cm} &  \sum_{i\in\V} J_i(\vec u_i; \vec x_{i, 0}) 
		\label{eq:probl:cost}\\
		\st \hspace{0.2cm} & \vec{\dot x}_i = \vec f_i(\vec x_i, \vec u_i, t) + \sum_{j\in\N_i^\sending} \vec f_{ij}(\vec x_i, \vec u_i, \vec x_j, \vec u_j, t),\quad i\in\V
		\label{eq:probl:basic_OCP_dynamics} \\
		& \vec x_i(0) = \vec x_{i, 0},\quad i \in \V
		\label{eq:probl:basic_OCP_dynamics_x0} \\
		& \vec 0 = \vec g_i(\vec x_i, \vec u_i, t),\quad i\in\V 
		\label{eq:probl:basic_OCP_const1}\\
		& \vec 0 = \vec g_{ij}(\vec x_i, \vec u_i, \vec x_j, \vec u_j, t),\quad j\in \N_i^\sending,\ i\in\V 
		\label{eq:probl:basic_OCP_const12}\\
		& \vec 0 \geq \vec h_i(\vec x_i, \vec u_i, t),\quad i\in\V 
		\label{eq:probl:basic_OCP_const13}\\
		& \vec 0 \geq \vec h_{ij}(\vec x_i, \vec u_i, \vec
           x_j, \vec u_j, t),\quad j\in \N_i^\sending,\ i\in\V 
           \label{eq:probl:basic_OCP_const14}\\ 
          \label{eq:probl:basic_OCP_constr_u}
		& \vec u_i \in \left[ \vec u_{i, \mathrm{min}},\ \vec u_{i, \mathrm{max}} \right],\quad i\in\V
	\end{align}
	\label{eq:probl:basic_OCP}%
\end{subequations}
with
\begin{align}
	J_i(\vec u_i; \vec x_{i, 0}) =V_i(\vec x_i(T), T) + \int_0^T l_i(\vec x_i, \vec u_i, t) \dt,
	\label{eq:global_cost}
\end{align}
states $\vec x_i(t)\in\R^{n_{x, i}}$, controls $\vec
u_i(t)\in\R^{n_{u, i}}$ and the horizon length
$T>0$.
Each agent may have a general nonlinear cost function $l_i\ :\
\R^{n_{x, i}}\times\R^{n_{u, i}}\times\R\rightarrow \R$ and terminal
cost $V_i\ : \ \R^{n_{x, i}}\times\R\rightarrow\R$. The overall cost
function \eqref{eq:probl:cost} is given by the sum over the individual cost functions. The
subsystem dynamics \eqref{eq:probl:basic_OCP_dynamics} are defined by the functions $\vec f_i\ : \ \R^{n_{x,
    i}}\times\R^{n_{u, i}}\times\R\rightarrow\R^{n_{x, i}}$ and $\vec
f_{ij}\ :\ \R^{n_{x, i}}\times\R^{n_{u, i}}\times\R^{n_{x,
    j}}\times\R^{n_{u, j}}\times\R\rightarrow\R^{n_{x, i}}$. The
OCP additionally considers 
nonlinear equality constraints~\eqref{eq:probl:basic_OCP_const1}-\eqref{eq:probl:basic_OCP_const12} and inequality constraints~\eqref{eq:probl:basic_OCP_const13}-\eqref{eq:probl:basic_OCP_const14} with the functions $\vec g_i\ :\ \R^{n_{x,
    i}}\times\R^{n_{u, i}}\times\R\rightarrow\R^{n_{g, i}}$, $\vec
g_{ij}\ :\ \R^{n_{x, i}}\times\R^{n_{u, i}}\times\R^{n_{x,
    j}}\times\R^{n_{u, j}}\times\R\rightarrow\R^{n_{g, ij}}$ and
$\vec h_i\ :\ \R^{n_{x, i}}\times\R^{n_{u,
    i}}\times\R\rightarrow\R^{n_{h, i}}$, $\vec h_{ij}\ :\ \R^{n_{x,
    i}}\times\R^{n_{u, i}}\times\R^{n_{x, j}}\times\R^{n_{u,
    j}}\times\R\rightarrow\R^{n_{h, ij}}$ as well as box
constraints~\eqref{eq:probl:basic_OCP_constr_u} for the control input $\vec u_i$ of each agent
$i\in\V$.
% may be bound between $\vec
%u_{i,\text{min}}\in\R^{n_{u, i}}$ and $\vec
%u_{i,\text{max}}\in\R^{n_{u, i}}$. 

The neighborhood $\N_i$ of agent $i\in\V$ is given by two sets that differ
in the direction of the coupling, \textit{sending neighbors}
$\N_i^\sending$ and \textit{receiving neighbors}
$\N_i^\receiving$, see Figure \ref{img:neighborhood} for an example. 
States and controls of sending neighbors have an 
explicit influence on the dynamics of the agent $i$ in form of
functions $\vec f_{ij}$, see
\eqref{eq:probl:basic_OCP_dynamics}. Receiving neighbors are neighbors
of agent $i$ that are explicitly influenced by this agent, hence
states or controls of the agent are part of a function $\vec f_{ij}$ of
receiving neighbors. While a neighbor can be both, receiving and
sending, this separation is going to be beneficial in the \ADMM\
algorithm by reducing unnecessary computation and communication
effort.  

The dynamics \eqref{eq:probl:basic_OCP_dynamics} of each agent $i\in\V$ are \textit{neighbor affine} in
the sense that the dynamics consists of a function $\vec
f_i$ that depend only on states and controls of the 
agent and a sum of functions  $\vec f_{ij}$ that
depend on states and controls of the agent and one neighbor. The constraints~\eqref{eq:probl:basic_OCP_const1}-\eqref{eq:probl:basic_OCP_constr_u}
on each agent are separated into
constraints that depend on states and controls of the agent, given by $\vec g_i$, $\vec h_i$ and the box constraints~\eqref{eq:probl:basic_OCP_constr_u}, and constraints
$\vec g_{ij}$ and $\vec h_{ij}$ depending on states and controls of the agent and one neighbor, similar
to the dynamics. 
%Agent constraint functions $\vec g_i$ and $\vec
%h_i$ only depend on states and controls of the agent $i$ while
%coupling constraints $\vec g_{ij}$ and $\vec h_{ij}$ may depend on
%states and controls of the agent and at most one neighbor. 

The considered OCP formulation \eqref{eq:probl:basic_OCP} covers a wide class of distributed
systems, e.g. cooperative transport \cite{bib:hentzelt2013} and scalable systems such as smart
grids \cite{bib:filatrella2008}. This generic system description combined
with the focus on a time-efficient implementation opens a wide
spectrum of usability for the presented DMPC framework.

\section{Distributed model predictive control}
\label{sec:DMPC}

Optimal control problems for coupled systems as in 
\eqref{eq:probl:basic_OCP} contain a large number of states and
controls. This leads to a significant computational effort that is
challenging for standard \MPC algorithms to be handled in real-time.
\DMPC algorithms instead assume that each of the distributed
subsystems are equipped with a dedicated control unit that is capable
of solving a reduced optimal control problem. The idea 
based on this assumption is to decouple the global OCP and spread the
computation effort over the set of agents in parallel. 
Overlying algorithms ensure convergence of the
local solutions to an optimal solution for the overall system. While
the computational complexity and communication effort of algorithms
for DMPC is higher than solving the central problem, the expectation
is to compensate this disadvantage by the parallel structure. In the
presented DMPC framework, the well-known \ADMM\ algorithm \cite{bib:boyd2011} is employed
in a continuous-time setting~\cite{bib:bestler2017}.
% following the idea of real-time
%MPC as it provides a fast convergence 
%to a sufficient precise solution \cite{bib:graichen2010}. 

%\input{src/4_1_AL}

\subsection{ADMM algorithm}

The \ADMM\ algorithm enables to spread the computation effort of the
global OCP \eqref{eq:probl:basic_OCP} completely on distributed
agents. %To derive the algorithm, 
As a starting point, the global OCP
\eqref{eq:probl:basic_OCP} is brought into a decoupled form for each
agent $i\in\V$ by introducing local copies $\vec{\bar x}_{ji}(t)\in\R^{n_{x,
    j}}$ and $\vec{\bar u}_{ji}(t)\in\R^{n_{u, j}}$ for the states
$\vec x_j$ and controls $\vec u_j$ of each sending neighbor
$j\in\N_i^\sending$, i.e.
\begin{subequations}
	\begin{align}
		\min_{\vec w, \vec z} \hspace{0.2cm} & \sum_{i\in\V} J_i(\vec u_i; \vec x_{i, 0}) \\
		\st \hspace{0.1cm} & \vec{\dot x}_i = \vec f_i(\vec x_i, \vec u_i, t) + \sum_{j\in\N_i^\sending} \vec f_{ij}(\vec x_i, \vec u_i, \vec{\bar x}_{ji}, \vec{\bar u}_{ji}, t),\quad i\in\V
		\label{eq:decoupled_dynamics}\\
		& \vec x_i(0) = \vec x_{i,0},\ i\in\V \\
		& \vec 0 = \vec g_i(\vec x_i, \vec u_i, t),\quad i\in\V \\
		& \vec 0 \geq \vec h_i(\vec x_i, \vec u_i, t),\quad i\in\V \\
		& \vec 0 = \vec g_{ij}(\vec x_i, \vec u_i, \vec{\bar x}_{ji}, \vec{\bar u}_{ji}, t),\quad j\in\N_i^\sending,\ i\in\V \\
		& \vec 0 \geq \vec h_{ij}(\vec x_i, \vec u_i, \vec{\bar x}_{ji}, \vec{\bar u}_{ji}, t),\quad j\in\N_i^\sending,\ i\in\V \\
		& \vec u_i \in \left[ \vec u_{i, \mathrm{min}},\ \vec u_{i, \mathrm{max}} \right],\quad i\in\V 
		\label{eq:decoupled_boxconstraints}\\
		& \vec 0 = \begin{bmatrix} \vec z_{x, i} \\ \vec z_{u, i} \end{bmatrix} - \begin{bmatrix} \vec x_i \\ \vec u_i \end{bmatrix},\quad i\in\V 
		\label{eq:decoupled_const_1}\\
		& \vec 0 = \begin{bmatrix} \vec z_{x, j} \\ \vec z_{u, j} \end{bmatrix} - \begin{bmatrix} \vec{\bar x}_{ji} \\ \vec{\bar u}_{ji} \end{bmatrix},\quad j\in\N_i^\sending,\ i\in\V.
		\label{eq:decoupled_const_2}
	\end{align}
	\label{eq:decoupled_OCP}%
\end{subequations}
%
%with decoupled dynamics and constraints. This is achieved 
These local copies $(\vec{\bar x}_{ji},\vec{\bar u}_{ji})$ represent
new control inputs for the agent $i$ and can be seen as a
proposal of agent $i$ for its neighbors $j\in\N_i^\sending$. 
%They are signed using a bar over the variable name with an index notation as
%follows. The first index indicates the agent that owns the original
%variable while the second index points to the agent that has the local
%copy, henc $\vec{\bar x}_{ji}$ is a local copy of the original state
%$\vec x_j$ of agent $j$ while the copy belongs to agent $i$. 
Equivalence between the local copies and the original variables is
ensured by introducing
%for the stationary solution, 
the consistency constraints
\eqref{eq:decoupled_const_1} and \eqref{eq:decoupled_const_2} with
the coupling variables $\vec z_{x,
  i}(t)\in\R^{n_{x, i}}$ and $\vec z_{u, i}(t)\in\R^{n_{u,
    i}}$. 
%The coupling variables $(\vec z_{x,i}, \vec z_{u,i})$ are a
%degree of freedom analogous to the local copies. 
In \eqref{eq:decoupled_OCP} and the following, the
notation
\begin{subequations}
	\begin{align}
		\vec w_i=&\ \begin{bmatrix} \vec u_i \\ \begin{bmatrix} \vec{\bar x}_{ji} \\ \vec{\bar u}_{ji} \end{bmatrix}_{j\in\N_i^\sending} \end{bmatrix}, \quad 
		\vec z_i = \begin{bmatrix} \vec z_{x, i} \\ \vec z_{u, i} \end{bmatrix}, \quad
		\vec z_{\shortminus i} = \begin{bmatrix} \vec z_{x, j} \\ \vec z_{u, j} \end{bmatrix}_{j\in\N_i^\sending},\quad 
		i\in\V \\ 
		\vec w =&\ \begin{bmatrix} \vec w_i \end{bmatrix}_{i\in\V},\quad \vec z = \begin{bmatrix} \vec z_i \end{bmatrix}_{i\in\V}
	\end{align}
	\label{eq:stacked_notation0}
\end{subequations}
is used. 
%Note the usage of $\vec C$ instead of $\vec\Rho$ regarding
%the diagonal matrices to avoid confusion with the weighting matrices
%or the terminating cost in Section \ref{sec:simulation_example}. 

The ADMM method is based on the Augmented Lagrangian formulation
\cite{bib:bestler2017,bib:bertsekas1996}. Regarding the continuous-time setting used
in this paper, the consistency constraints~\eqref{eq:decoupled_const_1} and \eqref{eq:decoupled_const_2} are
  accounted for in the cost functional
\begin{align}
	\begin{split}
		&J_{\rho, i}(\vec u_i, \vec \mu_i, \vec z_i, \vec z_{\shortminus i}; \vec x_{i, 0}) \\
		&\ = J_i(\vec u_i; \vec x_{i, 0}) \\
		&\ + \int_0^T  \begin{bmatrix} \vec \mu_{x, ii} \\ \vec \mu_{u, ii} \end{bmatrix}^\TT \hspace{-0.15cm}\left( \begin{bmatrix} \vec z_{x, i} \\ \vec z_{u, i} \end{bmatrix} - \begin{bmatrix} \vec x_i \\ \vec u_i \end{bmatrix} \right) 
		+ \frac{1}{2} \norm{ \begin{bmatrix} \vec z_{x, i} \\ \vec z_{u, i} \end{bmatrix} - \begin{bmatrix} \vec x_i \\ \vec u_i \end{bmatrix} }^2_{\vec C_i}  \\
		&\ + \sum_{j\in\N_i^\sending} \begin{bmatrix} \vec \mu_{x, ji} \\ \vec \mu_{u, ji} \end{bmatrix}^\TT \left( \begin{bmatrix} \vec z_{x, j} \\ \vec z_{u, j} \end{bmatrix} - \begin{bmatrix} \vec{\bar x}_{ji} \\ \vec{\bar u}_{ji} \end{bmatrix} \right) \hspace{-0.05cm}
		+ \frac{1}{2} \norm{ \begin{bmatrix} \vec z_{x, j} \\ \vec z_{u, j} \end{bmatrix} - \begin{bmatrix} \vec{\bar x}_{ji} \\ \vec{\bar u}_{ji} \end{bmatrix} }^2_{\vec C_{ji}}
		\dt
	\end{split}
	\label{eq:J_rho}%
\end{align}
subject to \eqref{eq:decoupled_dynamics}-\eqref{eq:decoupled_boxconstraints} with the Lagrange multipliers $\vec \mu_{x, ii}(t)\in\R^{n_{x,
		i}}$, $\vec \mu_{u, ii}(t)\in\R^{n_{u, i}}$, $\vec \mu_{x,
	ji}(t)\in\R^{n_{x, j}}$, $\vec \mu_{u, ji}(t)\in\R^{n_{u, j}}$
and penalty parameters $\vec \rho_{x, i}(t)\in\R^{n_{x, i}}$, $\vec
\rho_{u, i}(t)\in\R^{n_{u, i}}$, $\vec \rho_{x,
	ji}(t)\in\R^{n_{x, j}}$ and $\vec \rho_{u, ji}(t)\in\R^{n_{u,
		j}}$. To ease notations, the multipliers and penalty parameters are stacked according to
\begin{subequations}
	\begin{align}
		\vec \mu_i =&\ \begin{bmatrix} \vec \mu_{x, ii} \\ \vec \mu_{u, ii} \\ \begin{bmatrix} \vec \mu_{x, ji} \\ \vec \mu_{u, ji} \end{bmatrix}_{j\in\N_i^\sending} \end{bmatrix},\ i\in\V ,& 
		\vec \mu =&\ \begin{bmatrix} \vec \mu_i \end{bmatrix}_{i\in\V} \\
		\vec C_i =&\ \mathrm{diag}\begin{bmatrix} \vec \rho_{x, i} \\ \vec \rho_{u, i} \end{bmatrix},\ i\in\V &  
		\vec C_{ji} =&\  \mathrm{diag}\begin{bmatrix} \vec \rho_{x, ji} \\ \vec \rho_{u, ji} \end{bmatrix},\ {j\in\N_i^\sending},\ i\in\V.
	\end{align}
	\label{eq:stacked_notation1}
\end{subequations}
The corresponding dual problem to~\eqref{eq:decoupled_OCP} can be written as 
\begin{subequations}
	\begin{align}
	\max_{\vec \mu}\min_{\vec w, \vec z} \hspace{0.2cm} & \sum_{i\in\V} J_{\rho, i}(\vec u_i, \vec \mu_i, \vec z_i, \vec z_{\shortminus i}; \vec x_{i, 0}) \\
	\st & \vec{\dot x}_i = \vec f_i(\vec x_i, \vec u_i, \tau) + \sum_{j\in\N_i^\sending} \vec f_{ij}(\vec x_i, \vec u_i, \vec{\bar x}_{ji}, \vec{\bar u}_{ji}, t),\ i\in\V 
	\label{eq:OCP_dynamics}\\
	& \vec x_i(0) = \vec x_{i,0},\quad i\in\V 
	\label{eq:OCP_initial}\\
	& \vec 0 = \vec g_i(\vec x_i, \vec u_i, t),\quad i\in\V 
	\label{eq:OCP_constraint_1}\\
	& \vec 0 \geq \vec h_i(\vec x_i, \vec u_i, t),\quad i\in\V \\
	& \vec 0 = \vec g_{ij}(\vec x_i, \vec u_i, \vec{\bar x}_{ji}, \vec{\bar u}_{ji}, t),\quad j\in\N_i^\sending,\ i\in\V \\
	& \vec 0 \geq \vec h_{ij}(\vec x_i, \vec u_i, \vec{\bar x}_{ji}, \vec{\bar u}_{ji}, t),\quad j\in\N_i^\sending,\ i\in\V\\
	& \vec u_i \in \left[ \vec u_{i, \mathrm{min}},\ \vec u_{i, \mathrm{max}} \right],\quad i\in\V
	\label{eq:OCP_constraint_2}
	\end{align}
	\label{eq:resulting_OCP}%
\end{subequations}
with the primal variables $(\vec w,\vec z)$ and the dual
  variables $\vec\mu$.
The \ADMM\ algorithm solves the \textrm{max}-\textrm{min}-problem \eqref{eq:resulting_OCP} by repetitively executing the three steps 
\begin{subequations}
	\begin{align}
		\begin{split}
			\min_{\vec w} \hspace{0.2cm} & \sum_{i\in\V} J_{\rho, i}(\vec u_i, \vec \mu_i^{q-1}, \vec z_i^{q-1}, \vec z_{\shortminus i}^{q-1}; \vec x_{i, 0}),\quad \st\ \eqref{eq:OCP_dynamics} - \eqref{eq:OCP_constraint_2}
		\end{split} \\
		\min_{\vec z} \hspace{0.2cm} & \sum_{i\in\V} J_{\rho, i}(\vec u_i^q, \vec \mu_i^{q-1}, \vec z_i, \vec z_{\shortminus i}; \vec x_{i, 0})
		\label{eq:ADMMSteps_z} \\
		\vec \mu_i^q =&\ \vec \mu_i^{q-1} + \mathrm{diag}\begin{bmatrix} \vec C_i \\ \begin{bmatrix} \vec C_{ji} \end{bmatrix}_{j\in\N_i^\sending} \end{bmatrix} \begin{bmatrix} \vec z_i^q - \begin{bmatrix} \vec x_i^q \\ \vec u_i^q \end{bmatrix} \\ \begin{bmatrix} \vec z_j^q - \begin{bmatrix} \vec{\bar x}_{ji}^q \\ \vec{\bar u}_{ji}^q \end{bmatrix} \end{bmatrix}_{j\in\N_i^\sending} \end{bmatrix},\quad i\in\V
		\label{eq:steepest_ascent}
	\end{align}
\end{subequations}
with the iteration counter $q$. The minimization with respect to the coupling variables $\vec z$ \eqref{eq:ADMMSteps_z} can be solved analytically while the steepest ascent is used in \eqref{eq:steepest_ascent}. Important to note is that each step
can be subdivided into fully decoupled steps for either agent
$i\in\V$. Hence, the algorithm is fully distributable which allows to
spread the computation effort over all agents.  

\begin{algorithm}
	\caption{Alternating direction method of multipliers}
	\label{alg:ADMM}
	Initialize $\vec w_i^0$, $\vec z_i^0$, $\vec \mu_i^0$, choose $\vec C_i$, $\vec C_{ji}$, $\epsilon$, set $q=1$
	\begin{algorithmic}[1]
		\STATE Compute local variables $\vec w_i^{q}$ by solving
			\begin{subequations}
				\begin{align}
					\begin{split}
						\min_{\vec w_i} \hspace{0.2cm} & V_i(\vec x_i(T), T) + \hspace{-0.1cm} \int_0^T \hspace{-0.15cm} l_i(\vec x_i, \vec u_i, t) + \begin{bmatrix} \vec \mu_{x, ii}^{q-1} \\ \vec \mu_{u, ii}^{q-1} \end{bmatrix}^\TT \hspace{-0.2cm} \left( \vec z_i^{q-1}\text - \begin{bmatrix} \vec x_i \\ \vec u_i \end{bmatrix} \right) + \frac{1}{2} \norm{\vec z_i^{q-1}\text - \begin{bmatrix} \vec x_i \\ \vec u_i \end{bmatrix}}^2_{\vec C_i} \\
						&\ + \sum_{j\in\N_i^\sending} \begin{bmatrix} \vec \mu_{x, ji}^{q-1} \\ \vec \mu_{u, ji}^{q-1} \end{bmatrix}^\TT \left( \vec z_j^{q-1}\text - \begin{bmatrix} \vec{\bar x}_{ji} \\ \vec{\bar u}_{ji} \end{bmatrix} \right) + \frac{1}{2} \norm{ \vec z_j^{q-1}\text - \begin{bmatrix} \vec{\bar x}_{ji} \\ \vec{\bar u}_{ji} \end{bmatrix} }^2_{\vec C_{ji}} \dt
					\end{split} \\
					\st & \vec{\dot x}_i = \vec f_i(\vec x_i, \vec u_i, t) + \sum_{j\in\N_i^\sending} \vec f_{ij}(\vec x_i, \vec u_i, \vec{\bar x}_{ji}, \vec{\bar u}_{ji}, t),\quad \vec x_i(0) = \vec x_{i, 0} \\
					& \vec 0 = \vec g_i(\vec x_i, \vec u_i, t),\ \vec 0 = \vec g_{ij}(\vec x_i, \vec u_i, \vec{\bar x}_{ji}, \vec{\bar u}_{ji}, t),\quad j\in\N_i^\sending \\
					& \vec 0 \geq \vec h_i(\vec x_i, \vec u_i, t),\ \vec 0 \geq \vec h_{ij}(\vec x_i, \vec u_i, \vec{\bar x}_{ji}, \vec{\bar u}_{ji}, t),\quad j\in\N_i^\sending \\
					& \vec u_i \in \left[ \vec u_{i, \mathrm{min}},\ \vec u_{i, \mathrm{max}} \right] \label{eq:set}
				\end{align}
				\label{eq:ADMM_step1}
			\end{subequations}
		\STATE Send local copies $\vec{\bar x}_{ji}^q$ and $\vec{\bar u}_{ji}^q$ to sending neighbors $j\in\N_i^\sending$ \\
		\STATE Compute coupling variables 
			\begin{align}
				\vec z_i^q =&\ \frac{1}{1 + \abs{ \N_i^\receiving }} \left( \begin{bmatrix} \vec x_i^q \\ \vec u_i^q \end{bmatrix} - \vec C_i^{-1} \begin{bmatrix} \vec \mu_{x, ii}^{q-1} \\ \vec \mu_{u, ii}^{q-1} \end{bmatrix} + \sum_{j\in\N_i^\receiving} \begin{bmatrix} \vec{\bar x}_{ij}^q \\ \vec{\bar u}_{ij}^q \end{bmatrix} - \vec C_{ij}^{-1} \begin{bmatrix} \vec \mu_{x, ij}^{q-1} \\ \vec \mu_{u, ij}^{q-1} \end{bmatrix} \right)
				\label{eq:ADMM_step3}
			\end{align}
		\STATE Send coupling variables $\vec z_i^q$ to receiving neighbors $j\in\N_i^\receiving$
		\STATE Compute Lagrange multipliers 
			\begin{align}
				\vec \mu_i^q =&\ \vec \mu_i^{q-1} + \diag{\vec C_i, \vec C_{ji}} \begin{bmatrix} \vec z_i^q - \begin{bmatrix} \vec x_i^q \\ \vec u_i^q \end{bmatrix} \\ \begin{bmatrix} \vec z_j^q - \begin{bmatrix} \vec{\bar x}_{ji}^q \\ \vec{\bar u}_{ji}^q \end{bmatrix} \end{bmatrix}_{j\in\N_i\sending} \end{bmatrix}
				\label{eq:ADMM_step5}
			\end{align}
		\STATE Send Lagrange multipliers $\vec \mu_i^q$ to sending neighbors $j\in\N_i^\sending$ \\
		\IF {if $q=q_{\mathrm{max}}$ or $\norm{ \begin{bmatrix} \vec z_i^q - \vec z_i^{q-1} \\ \vec \mu_i^q - \vec \mu_i^{q-1} \end{bmatrix} } \leq \epsilon,\ \forall\ i\in\V$}
			\STATE STOP
		\ELSE
			\STATE
			set $q=q+1$ and go to Step 1.
		\ENDIF
	\end{algorithmic}
\end{algorithm}

The resulting \ADMM\ algorithm for each agent is given in Algorithm
\ref{alg:ADMM}. It consists of the computation steps $1$, $3$, $5$,
the communication steps $2$, $4$, $6$, and the evaluation of a
convergence criterion in Step 7. The algorithm starts with an
initialization of corresponding variables. The local OCP
\eqref{eq:ADMM_step1} is minimized in Step 1 with respect to the local
variables $\vec w_i$. This minimization represents the main computation
effort of the overall algorithm. In Step 2, the trajectories of the local
variables are sent to the sending neighbors
$j\in\N_i^\sending$ of each agent $i\in\V$. The analytic solution for
the minimization with respect to the coupling variables~\eqref{eq:ADMMSteps_z} is given in
Step 3 of the \ADMM\ algorithm, before they are sent to the
receiving neighbors $j\in\N_i^\receiving$ of each agent in Step 4. The
third computation step is given in Step 5 by a maximization with
respect to the Lagrange multipliers $\vec\mu$. In Step 6, the result of the
maximization step is sent to the sending neighbors
$j\in\N_i^\sending$ of each agent $i\in\V$. A convergence criterion is
checked in Step 7. If it is satisfied or the iteration counter has
reached its maximum, the algorithm stops and returns the current
trajectories. Otherwise, the iteration counter is increased and the
algorithm returns to Step 1.  

\subsection{Neighbor approximation}
\label{subsec:neighbor_approx}

In practice, the convergence speed of the ADMM algorithm can be enhanced by
anticipating the actions of the neighbors in the own agents
optimization. The concept of neighbor approximation was introduced in
\cite{bib:hentzelt2013} and extended in \cite{bib:burk2020} and relies
on the neighbor affine structure of the dynamics~\eqref{eq:probl:basic_OCP_dynamics}-\eqref{eq:probl:basic_OCP_dynamics_x0} and
  constraints~\eqref{eq:probl:basic_OCP_const1}-\eqref{eq:probl:basic_OCP_constr_u}. 
%The concept of neighbor approximation can be used to improve the
%convergence behavior of the \ADMM\ algorithm. It was introduced in
%\cite{bib:hentzelt2013} and extended in \cite{bib:burk2020}. 

The basic idea is to use the already introduced local copies $\vec{\bar{x}}_{ji}$ and $\vec{\bar{u}}_{ji}$ 
to approximate parts of the neighbors OCP. The expectation is that the
additional 
information about the neighborhood improves the local solution of each
agent and thus the convergence behavior of the overall
algorithm. This also reduces the number of required \ADMM\
iterations until convergence is reached, which has been confirmed in
numerical evaluations in \cite{bib:hentzelt2013} and
\cite{bib:burk2020}. 
In practical experience, 
%the combination of fewer \ADMM\
%iterations with the relaxed parameterization of the underlying solver
the reduced number of \ADMM iterations can compensate for the increased complexity of the extended OCP which can lead to a significantly 
decreased computational effort~\cite{bib:burk2020}. 
%See \cite{bib:burk2020} for a
%comprehensive numerical evaluation. 

The neighbor approximation implemented in \GRAMPCD is modular in
the sense that the neighbors cost, constraints, dynamics and each
combination of the three can be considered. 
%Hence, each approximation method is
%presented in the following separated from each other. 
%
%Note that local
%copies have to be introduced for receiving neighbors as well if
%neighbor approximation is used as in the presented DMPC-framework the
%OCP of both receiving and sending neighbors is approximated. 

\subsubsection{Neighbor cost}

The global cost to be mininized~\eqref{eq:global_cost} consists of the single cost
functions of the agents $i\in\V$. The local copies of the neighbor
variables $\vec{\bar{x}}_{ji}$ and $\vec{\bar{u}}_{ji}$, $j\in\N_i$, can be used to anticipate the
neighbors cost $J_j(\vec{\bar u}_{ji}; \vec x_{j, 0})$ on the local
level of agent $i\in\V$, i.e.\
%
%Approximating the neighbors cost function is presented as a second
%part of neighbor approximation. Therefore, note that the concept of
%neighbor approximation must not change either the optimal solution of
%the global OCP nor the absolute value of the cost but only affect the
%convergence behavior. If the neighbors cost function is approximated
%by simply including it into the local OCP in analogy to the
%approximation of constraints 
%
\begin{align}
		\tilde J_i(\vec u_i; \vec x_{i, 0}) =&\ \eta_i J_i(\vec u_i; \vec x_{i, 0}) + \sum_{j\in\N_i} \eta_j J_j(\vec{\bar u}_{ji}; \vec x_{j, 0}),\quad i\in\V\,.
\end{align} 
The normalization with the factors
%
%the neighbors cost function would appear in the overall cost function
%multiple times and the optimal solution would be affected. Hence,
%normalization factors  
\begin{align}
	\eta_i =&\ \frac{1}{1 + \abs{\N_i}},\quad i\in\V
\end{align}
is necessary in order to avoid that the neighbors cost function would
appear in the overall cost function multiple times.
% ensures that the  
%have to be introduced. They ensure the equivalence of the absolute
%value of the cost of the overall OCP with and without approximation of
%the neighbors cost for a stationary solution. 
Approximating the neighbor costs is especially beneficial in examples
with a strong dependency on the other agents costs and enables the agent to
anticipate the neighbors control action to minimize its local costs.
It is recommended to combine the neighbor cost
approximation with the approximation of the neighbor dynamics
introduced in the following lines.
%function with the approximation of neighbors dynamics.
%While neighbor approximation is a modular
%concept, it is recommended to combine approximating neighbors cost
%function with the approximation of neighbors dynamics.

\subsubsection{Neighbor dynamics}

Similar to the neighbor cost consideration, the neighbor affine
structure of the dynamics~\eqref{eq:probl:basic_OCP_dynamics}-\eqref{eq:probl:basic_OCP_dynamics_x0} can be exploited to approximate
the neighbor dynamics and therefore to improve the quality of the
local copies $\vec{\bar{x}}_{ji}$ and $\vec{\bar{u}}_{ji}$. 
To this end, the local dynamics~\eqref{eq:probl:basic_OCP_dynamics}-\eqref{eq:probl:basic_OCP_dynamics_x0} are extended by the 
approximate neighbor dynamics 
%The third part of neighbor approximation is given by approximating the
%neighbors dynamics. Therefore, differential equations for the local
%copies  
%
\begin{align}
		&\vec{\dot{\bar x}}_{ji} = \vec f_j(\vec{\bar x}_{ji}, \vec{\bar u}_{ji}, t) + \vec f_{ji}(\vec{\bar x}_{ji}, \vec{\bar u}_{ji}, \vec x_i, \vec u_i, t) 
		 + \vec{\bar v}_{ji},\quad j\in\N_i,\ i\in\V
		 %+\hspace{-0.3cm} \underbrace{\sum_{s\in\N_j^\sending\setminus i}\hspace{-0.1cm} \vec f_{js}(\vec x_j, \vec u_j, \vec x_s, \vec u_s, t)}_{\vec{\bar v}_{ji}}
		 \label{eq:approx_dyn}%
\end{align}
with the initial condition $\vec{\bar x}_{ji}(0) = \vec{ x}_{j, 0}$. The dependencies of the neighbor's states and controls in $\vec f_j$ and $\vec f_{ji}$ are
decoupled using the local copies $\vec{\bar{x}}_{ji}$ and $\vec{\bar{u}}_{ji}$. 
However, it is not possible to decouple further
functions $\vec f_{js}$ as these depend on states
and controls of agents $s$ for which agent $i$ has in general no local
copies. For consistency, the \textit{external influence}
\begin{align}
	\vec v_{ij} =&\ \sum_{s\in\N_i^\sending\setminus\{j\}} \vec
	f_{is}(\vec x_i, \vec u_i, \vec x_s, \vec u_s,
	t),\quad\ j\in\N_i,\ i\in\V \, 
\end{align}
is introduced with  $\vec v_{ij}(t)\in\R^{n_{x, i}}$ that captures the
remaining terms of the neighbors dynamics. The external influence is 
considered in the approximated neighbor dynamics~\eqref{eq:approx_dyn}
by introducing local copies $\vec{\bar v}_{ji}(t)\in\R^{n_{x, j}}$.
Thereby, the whole
dynamics of neighbor $j$ is approximated in \eqref{eq:approx_dyn}. To ensure convergence of the local copies $\vec{\bar v}_{ji}$ to the original variables $\vec v_{ij}$, the consistency constraints
\begin{subequations}
	\begin{align}
		\vec z_{v, ij} =&\ \vec v_{ij},\quad j\in\N_i,\ i\in\V \\
		\vec z_{v, ji} =&\ \vec{\bar v}_{ji},\quad j\in\N_i,\ i\in\V
	\end{align}
\end{subequations}
are introduced and replace the consistency constraints in~\eqref{eq:decoupled_const_1}-\eqref{eq:decoupled_const_2} regarding the states.
Note that the local copies of the states $\vec{\bar x}_{ji}$ are not considered
as control variables anymore, but are determined by the differential
equation \eqref{eq:approx_dyn}. Instead, the local copies of the
external influence $\vec{\bar v}_{ji}$ serve as new local control variables. 
In summary, the stacked notations~\eqref{eq:stacked_notation0} and~\eqref{eq:stacked_notation1} are adapted  according to
\begin{subequations}
	\begin{align}
			\vec w_i=&\ \begin{bmatrix} \vec u_i \\ \begin{bmatrix} \vec{\bar u}_{ji} \\ \vec{\bar v}_{ji} \end{bmatrix}_{j\in\N_i} \end{bmatrix},\ i\in\V& 
			\vec z_i =&\ \begin{bmatrix} \vec z_{u, i} \\ \begin{bmatrix} \vec z_{v, ij} \end{bmatrix}_{j\in\N_i} \end{bmatrix},\ i\in\V \\
			\vec \mu_i =&\ \begin{bmatrix} \vec \mu_{u, ii} \\ \begin{bmatrix} \vec \mu_{v, ij} \\ \vec \mu_{u, ji} \\ \vec \mu_{v, ji} \end{bmatrix}_{j\in\N_i} \end{bmatrix},\ i\in\V&
			\vec z_{\shortminus i} =&\ \begin{bmatrix} \vec z_{u, j} \\ \vec z_{v, ji} \end{bmatrix}_{j\in\N_i},\ i\in\V \\
			\vec C_i =&\ \mathrm{diag}\begin{bmatrix} \vec \rho_{u, i}  \end{bmatrix},\ i\in\V&
			\vec C_{ji} =&\ \mathrm{diag}\begin{bmatrix} \vec \rho_{v, ij} \\ \vec \rho_{u, ji} \\ \vec \rho_{v, ji} \end{bmatrix},\ {j\in\N_i},\ i\in\V 
	\end{align}
	\label{eq:new_notation}%
\end{subequations}
and
\begin{align}
	\vec w =&\ \begin{bmatrix} \vec w_i \end{bmatrix}_{i\in\V},& \vec z =&\ \begin{bmatrix} \vec z_i \end{bmatrix}_{i\in\V},& \vec \mu =&\ \begin{bmatrix} \vec \mu_i \end{bmatrix}_{i\in\V}
\end{align}
with Lagrangian multipliers $\vec \mu_{v, ij}(t)\in\R^{n_{x, i}}$,
$\vec\mu_{v, ji}(t)\in\R^{n_{x, j}}$, coupling variables $\vec
z_{v, ij}(t)\in\R^{n_{x, i}}$, and penalty parameters $\vec \rho_{v,
  ij}(t)\in\R^{n_{x, i}}$, $\vec \rho_{v, ji}(t)\in\R^{n_{x,
    j}}$.

\subsubsection{Neighbor constraints}%Approximate constraints}

In addition to the consideration of the neighbor cost and dynamics
within the local OCP of agent $i\in\V$, the constraints~\eqref{eq:probl:basic_OCP_const1}-\eqref{eq:probl:basic_OCP_constr_u} of each
neighbor $j\in\N_i$ of agent $i\in\V$ can be taken into account by adding 
%At first, the approximation of the neighbors constraints is
%presented. Therefore, the constraints of each neighbor $j\in\N_i$ 
%
\begin{subequations}
	\begin{align}
		& \vec 0 = \vec g_j(\vec{\bar x}_{ji}, \vec{\bar u}_{ji}, t),\quad \vec 0 = \vec g_{ij}(\vec{\bar x}_{ji}, \vec{\bar u}_{ji}, \vec x_i, \vec u_i, t),\quad j\in\N_i,\ i\in\V \\
		& \vec 0 \geq \vec h_i(\vec{\bar x}_{ji}, \vec{\bar u}_{ji}, t),\quad \vec 0 \geq \vec h_{ij}(\vec{\bar x}_{ji}, \vec{\bar u}_{ji}, \vec x_i, \vec u_i, t),\quad j\in\N_i,\ i\in\V \\
		& \vec{\bar u}_{ji} \in \left[ \vec u_{j, \mathrm{min}},\ \vec u_{j, \mathrm{max}} \right],\quad j\in\N_i,\ i\in\V \label{eq:approx_set}
	\end{align}
\end{subequations}
%
%are included into 
to the local OCP \eqref{eq:ADMM_step1}. % of agent $i\in\V$. 
Again, the constraints are decoupled from the neighbors states and
controls $\vec x_j$ and $\vec u_i$ by using the local copies $\vec{\bar x}_{ji}$ and $\vec{\bar u}_{ji}$.

As discussed before, this concept is restricted to
the agent constraints of each neighbor $j\in\N_i$ and the coupling
constraints between neighbors $j$ and agent $i$, while further
coupling constraints between neighbor $j$ and its neighbors
$s\in\N_j^\sending\setminus\{i\}$ depend on states and controls of
agents $s$ for which in general agent $i$ has no local copies. 
%Approximating the neighbors constraints on states and
%controls leads to constraints on the local copies of the agent and by
%this favors to generate feasible trajectories.  

\subsection{Penalty parameter adaption}
\label{subsec:penalty_adaption}

The update of the penalty parameters in the matrices $\vec C_i$ and $\vec C_{ji}$ in~\eqref{eq:resulting_OCP} is
crucial for a fast convergence of the \ADMM\ algorithm. 
%As a good choice may change from iteration to 
%iteration, adaption methods are required. 
The adaptation method implemented in \GRAMPCD follows 
%presented implementation in the presented framework follows 
a proposal in \cite[Section
3.4.1]{bib:boyd2011} for the optimization problem 
\begin{subequations}
	\begin{align}
		\min_{\vec x, \vec z} \hspace{0.2cm} & f(\vec x) + g(\vec z) \\
		\st \hspace{0.1cm} & \vec A\vec x + \vec B\vec z = \vec c\ .
	\end{align}
\end{subequations}
The proposed adaption algorithm is given by
\begin{align}
	\rho^{q} =
	\begin{cases}
		\tau^\mathrm{incr} \rho^{q-1} & \text{if } \norm{\vec r^{q-1}}_2 > \mu \norm{\vec s^{q-1}}_2 \\
			\frac{\rho^{q-1}}{\tau^\mathrm{decr}} & \text{if } \norm{\vec s^{q-1}}_2 > \mu \norm{\vec r^{q-1} }_2 \\
			\rho^{q-1} & \text{otherwise}
	\end{cases}
	\label{eq:adapt_boyd}
\end{align}
with the primal residual $\vec r^q = \vec A \vec x^q + \vec B \vec z^q
- \vec c$, the dual residual $\vec s^q = \rho \vec A^\TT \vec B (\vec
z^q - \vec z^{q-1})$. The basic idea is to keep both within a factor
of $\mu$ of one another. Following this idea for the OCP \eqref{eq:resulting_OCP},
the primal and dual residuals are given by 
\begin{subequations}
	\begin{align}
		\vec r^q_i =&\ \begin{bmatrix} \vec x_i^q \\ \Vec u_i^q \\ \begin{bmatrix} \vec{\bar x_{ji}}^q \\ \vec{\bar u}_{ji}^q \end{bmatrix}_{j\in\N_i^\sending} \end{bmatrix} - \begin{bmatrix} \vec z_{x, i}^q \\ \vec z_{u, i}^q \\ \begin{bmatrix} \vec z_{x, j}^q \\ \vec z_{u, j}^q \end{bmatrix}_{j\in\N_i^\sending} \end{bmatrix},\quad i\in\V \\
		\vec s^q_i =&\ \mathrm{diag}\begin{bmatrix} \vec C_i^{q-1} \\ \begin{bmatrix} \vec C_{ji}^{q-1} \end{bmatrix}_{j\in\N_i^\sending} \end{bmatrix} \begin{bmatrix} \vec z_{x, i}^q - \vec z_{x, i}^{q-1} \\ \vec z_{u, i}^q - \vec z_{u, i}^{q-1} \\ \begin{bmatrix} \vec z_{x, j}^q  - \vec z_{x, j}^{q-1}\\ \vec z_{u, j}^q  - \vec z_{u, j}^{q-1}\end{bmatrix}_{j\in\N_i^\sending} \end{bmatrix},\quad i\in\V.
	\end{align}
\end{subequations}
To reduce the 
number of tuning parameters, $\mu=1$ is chosen which results in an equality instead of the inequality in \eqref{eq:adapt_boyd}. To further simplify the implementation, the equality is evaluated element-wise and at each discrete time step $\delta_k = \frac{T}{N-1}$ with $N$ as discretization of the predicted horizon. This results in the condition
\begin{align}
	\norm{\vec r_i^q(\delta_k)}_2 \overset{!}{=} \norm{\vec s^q_i(\delta_k)}_2 
	\label{eq:rho:_update2}%
\end{align}
for each discrete time step $\delta_k$ and the norm evaluated element-wise. The condition~\eqref{eq:rho:_update2} can be reformulated in form of the update law 
\begin{align}
	\rho^q_m(\delta_k) =&\ \rho^{q-1}_m(\delta_k) \frac{\abs{r^q_m(\delta_k)}}{\abs{s^q_m(\delta_k)}} = \rho^{q-1}_m(\delta_k) \gamma^q_m(\delta_k)
\end{align}
with $m$ as index for an arbitrary element in \eqref{eq:rho:_update2}. The implementation is presented in Algorithm
\ref{alg:penaltyAdaption}. At first, the division through small
numbers, especially zero, is caught to prevent numerical issues. The
factor $\gamma^q$ is computed afterwards and bound between
$\gamma_\mathrm{min}$ and $\gamma_\mathrm{max}$, before the new penalty
parameter is calculated by $\rho^q = \gamma^q \rho^{q-1}$.  
\begin{algorithm}
	\caption{Adaption of penalty parameters}
	\label{alg:penaltyAdaption}
	Calculate $\abs{s^q}$ and $\abs{ r^q }$
	\begin{algorithmic}[1]
		\IF {$\abs{s^q} > \epsilon_0$}
			\STATE $\gamma^q = \frac{ \abs{ r^q } }{ \abs{ s^q } }$
			\IF{$\gamma^q > \gamma_\mathrm{max}$}
				\STATE $\gamma^q = \gamma_\mathrm{max}$
			\ELSIF{$\gamma^q < \gamma_\mathrm{min}$}
				\STATE $\gamma^q = \gamma_\mathrm{min}$
			\ENDIF
		\ELSE
			\STATE $\gamma^q = 1$
		\ENDIF
		\STATE $\rho^q = \gamma^q \rho^{q-1}$
	\end{algorithmic}
\end{algorithm}

\section{Modular framework}
\label{sec:framework}

\GRAMPCD is implemented in a modular fashion in order to achieve a scalable and flexible implementation.
%This section presents the modular framework. 
At first, the modular structure is explained before the capability for
plug-and-play scenarios is laid out.

\subsection{Modular structure}

The main parts of \GRAMPCD and their interaction are
presented in the following. Due to the modular concept, the
implementation of \GRAMPCD can be subdivided into single
modules that are composed depending on the chosen type of controller,
centralized or distributed, as the structure of \GRAMPCD
differs between the two cases. Both are visualized in Figure
\ref{img:CommunicationStructure}. Either structure is generated
automatically by choosing the corresponding controller type without
further required interaction of the user. Both the distributed and
centralized structure are scalable due to the modular concept and
therefore suitable to handle large or complex systems.

%Distributed
The distributed control structure in the left part of Figure
\ref{img:CommunicationStructure} assumes that each agent only has
access to its local variables and communication is required to acquire
data from other agents. 
Thus, the central part in the distributed setup is 
%Hence, the 
%of the distributed controller is a 
the communication interface. While it enables to exchange
data between agents, the actual implementation depends on the chosen
type of communication interface. If the \DMPC\ is simulated on a single
processor, there is no need to actually send data over a
network. Instead, a central communication interface is provided that
exchanges data pointers, which is a significant difference in
performance. 
%For this purpose, a central communication interface is provided. 
If the \ADMM\ algorithm is implemented in an actual
distributed setup, 
%the local communication 
%interface can be chosen. In this case, 
each agent creates its own
local communication interface that enables to exchange data over a
network. The corresponding protocol is encapsulated into the local
communication interface due to the modular concept that enables
implementing multiple protocols and switching between them. In either
case, each agent creates a local solver that contains the local OCP
depending on the neighborhood and the chosen optimization parameters
such as neighbor approximation. Hence, tasks like decoupling the
global OCP by introducing local copies are done automatically in the
background. The ADMM algorithm is implemented inside the local solver
with an abstract implementation of the minimization problem with
respect to the local variables. This enables implementing multiple
solvers and switching between them without changing other parts of the
software, although GRAMPC is chosen as default. The remaining two important modules are the coordinator and
the simulator. The \ADMM\ algorithm assumes a fully synchronized
execution, which has to be guaranteed even in a distributed setup. This
synchronization is handled by the coordinator by triggering each step
of the algorithm and waiting for a response of each agent before
sending the following trigger. The last module is an integrated
simulator that enables simulations independent of the chosen
controller or the specific system.

\newcommand{\versatz}{6.75}
\newcommand{\versatzUnten}{-1.75}
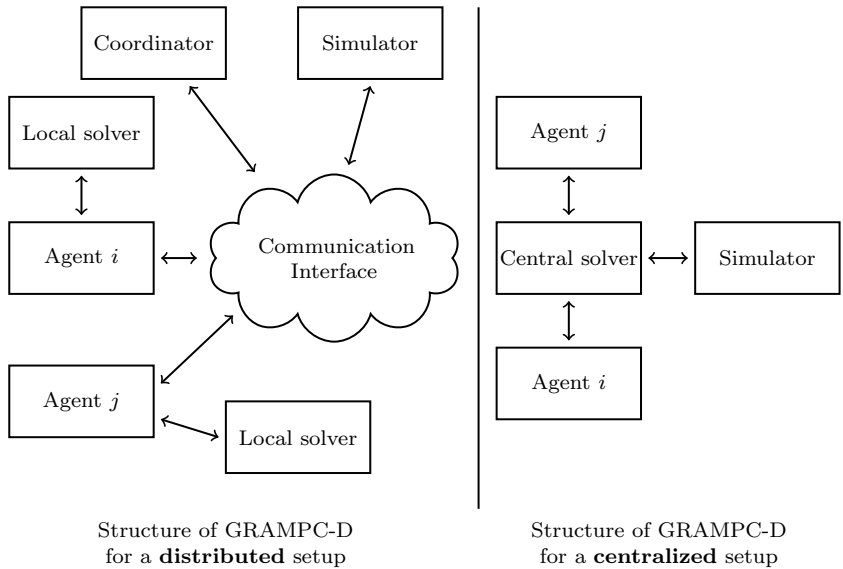
\begin{figure}[b]
	\centering
	\begin{tikzpicture}[scale=0.95]
		% Agent i
		\draw[line width = \linewidthfigure] (-1,0) rectangle (1, 1);
		\node[line width = \linewidthfigure] at (0, 0.5) [] {Agent $i$};
		\draw[line width = \linewidthfigure] (-1,1.75) rectangle (1, 2.75);
		\node[line width = \linewidthfigure] at (0, 2.25) [] {Local solver};
		\draw[line width = \linewidthfigure, <->] (1.1, 0.5) -- (1.6, 0.5);
		\draw[line width = \linewidthfigure, <->] (0, 1.1) -- (0, 1.65);
		% Agent j
		\draw[line width = \linewidthfigure] (-1,-1) rectangle (1, -2);
		\node[line width = \linewidthfigure] at (0, -1.5) [] {Agent $j$};
		\draw[line width = \linewidthfigure] (2,-1.5) rectangle (4, -2.5);
		\node at (3, -2) [] {Local solver};
		\draw[line width = \linewidthfigure, <->] (1.1, -1.25) -- (2.1, -0.3);
		\draw[line width = \linewidthfigure, <->] (1.1, -1.75) -- (1.9, -2);
		% Coordinator
		\draw[line width = \linewidthfigure] (0,3) rectangle (2, 4);
		\node at (1, 3.5) [] {Coordinator};
		\draw[line width = \linewidthfigure, <->] (1.5, 2.9) -- (2.4, 1.7);
		% Simulator
		\draw[line width = \linewidthfigure] (3, 3) rectangle (5, 4);
		\node at (4, 3.5) [] {Simulator};
		\draw[line width = \linewidthfigure, <->] (4, 2.9) -- (3.7, 1.8);
		% Communication
		\node[line width = \linewidthfigure, cloud, draw, aspect=2, text width = 2cm, align=center] at (3.5, 0.5) {Communication Interface};
		%%%%%%%%% big line
		\draw[line width = \linewidthfigure] (5.5, -3) -- (5.5, 4);
		%%%%%%%%% big line
		% Agent i
		\draw[line width = \linewidthfigure] (-1+\versatz,0+\versatzUnten) rectangle (1+\versatz, 1+\versatzUnten);
		\node[line width = \linewidthfigure] at (0+\versatz, 0.5+\versatzUnten) [] {Agent $i$};
		\draw[line width = \linewidthfigure] (-1+\versatz,1.75+\versatzUnten) rectangle (1+\versatz, 2.75+\versatzUnten);
		\node[line width = \linewidthfigure] at (0+\versatz, 2.25+\versatzUnten) [] {Central solver};
		\draw[line width = \linewidthfigure, <->] (0+\versatz, 1.1+\versatzUnten) -- (0+\versatz, 1.65+\versatzUnten);
		% Agent j
		\draw[line width = \linewidthfigure] (-1+\versatz,3.5+\versatzUnten) rectangle (1+\versatz, 4.5+\versatzUnten);
		\node[line width = \linewidthfigure] at (0+\versatz, 4+\versatzUnten) [] {Agent $j$};
		\draw[line width = \linewidthfigure, <->] (0+\versatz, 3.4+\versatzUnten) -- (0+\versatz, 2.85+\versatzUnten);
		% Simulator
		\draw[line width = \linewidthfigure] (1.75+\versatz, 1.75+\versatzUnten) rectangle (3.75+\versatz, 2.75+\versatzUnten);
		\node at (2.75+\versatz, 2.25+\versatzUnten) [] {Simulator};
		\draw[line width = \linewidthfigure, <->] (1.1+\versatz, 2.25+\versatzUnten) -- (1.65+\versatz, 2.25+\versatzUnten);
		%%%%%%%%%
		\node at (2, -3.5) [text width = 5cm, align=center] {Structure of \GRAMPCD for a \textbf{distributed} setup};
		\node at (8, -3.5) [text width = 5cm, align=center] {Structure of \GRAMPCD for a \textbf{centralized} setup};
	\end{tikzpicture}
	\caption{The communication interface is the central part of \GRAMPCD in case of distributed optimization. Each agent has its own local solver that handles the steps of the \ADMM\ algorithm. The coordinator provides a synchronization of all agents while the simulator handles the simulation of the overall system. If a centralized controller is chosen, \GRAMPCD is centered around the central solver that knows all agents and can access their variables.}
	\label{img:CommunicationStructure}
\end{figure}

%Centralized
A more simple structure is generated if a centralized controller is
chosen, see right part of Figure \ref{img:CommunicationStructure}. In
this case, the global OCP \eqref{eq:probl:basic_OCP} is solved in a centralized
manner 
%without the requirement for a communication
%interface. Instead, a centralized solver is generated 
including all agents dynamics and having knowledge of all
variables. The centralized setup implicitly synchronizes the execution
of the algorithm without the need for a coordinator. The only
remaining module is the simulator 
that is used to simulate the overall system.

%Exchange
Each part of \GRAMPCD is interchangeable by alternative software. 
%In version $1.0$ of the framework, 
As already mentioned, the MPC toolbox \GRAMPC\ is used by default to solve
both the global OCP~\eqref{eq:probl:basic_OCP} in case of a centralized controller and the
underlying reduced OCP~\eqref{eq:ADMM_step1} in Step $1$ of the \ADMM\ algorithm in case of
a distributed controller. \GRAMPC\ is tailored to embedded hardware
and by this a natural choice, but if another solver is desired, e.g.\
with a stronger focus on precision instead of computational speed, 
then the only required change is to overload the class regarding the
solver with a new implementation. The same holds for each part of the
framework such as the implemented communication protocol. 
The TCP protocol is provided by default, but alternative protocols
can be implemented by overloading the local communication interface.

%Real plants
%The modular concept enables additionally controlling actual plants with little changes to the software. The first required change is to apply the trajectories for the controls on the actual plant instead of the simulator and the second change to use measurements of the real system instead of states resulting from the simulation.

\subsection{Rapid prototyping}

\begin{figure}[b]
	\centering
	\begin{tikzpicture}
		%Probl. descr
		\draw[line width = \linewidthfigure] (0, 0) rectangle (3, 1);
		\node at (1.5, 0.5) [] {Problem description};
		% arrows
		\draw[line width = \linewidthfigure, ->] (1, -0.1) -- (-1, -1.9) node[midway,above, rotate=220.5+180]{Generate}node[midway,below, rotate=220.5+180]{central OCP};
		\draw[line width = \linewidthfigure, ->] (2, -0.1) -- (4, -1.9) node[midway,above, rotate=137.5+180]{Generate}node[midway,below, rotate=137.5+180]{local OCPs};
		% MPC
		\draw[line width = \linewidthfigure] (-2, -3) rectangle (0, -2);
		\node at (-1, -2.5) [] {MPC};
		% DMPC
		\draw[line width = \linewidthfigure] (2, -3) rectangle (6, -2);
		\node at (4, -2.5) {DMPC};
		\draw[line width = \linewidthfigure] (2, -4) rectangle (4, -3);
		\node at (3, -3.5) {Default};
		\draw[line width = \linewidthfigure] (6, -4) rectangle (4, -3);
		\node at (5, -3.5) {Neigh. approx.};
	\end{tikzpicture}
	\caption{Based on the same problem description, the global OCP
          is generated for a centralized controller and the local OCPs
          in the distributed case. The local OCPs are automatically
          extended by corresponding terms if neighbor approximation is
          enabled for the distributed controller.}
	\label{img:structure_ocp}
\end{figure}
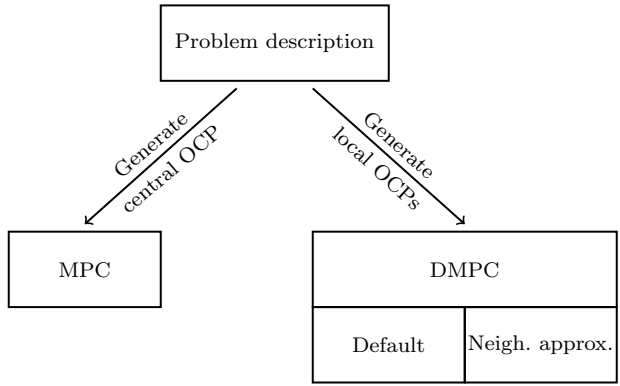

%In the presented DMPC-framework, b
Both the local and global OCPs are dynamically generated at
run-time based on the same problem description provided by the user,
see Figure \ref{img:structure_ocp}. In case of a centralized
controller, the global OCP is generated by the central solver while in
case of a distributed controller, the local OCP is generated for each
agent individually based on its neighbors and the optimization
parameters. Therein, the corresponding flags for neighbor
approximation are defined that state whether additional variables have
to be initialized, e.g. local copies $\vec{\bar x}_{ji}$ and $\vec{\bar u}_{ji}$ or the external
influence $\vec v_{ij}$, see
Section \ref{subsec:neighbor_approx}. This results in a convenient
prototyping process while designing controllers, as each type of
controller can be automatically generated and evaluated based on the same
problem description. To further support the efficiency of
prototyping, the possibility of multi-threading can be activated to
spread the computation effort on each available core of the processor
and thus to speed-up the computations.
%by this reduce the computation time significantly. 

\subsection{Plug-and-play functionality}

Plug-and-play is a core feature for the usability of a framework in
the field of \DMPC, 
%as the assumption of static problem descriptions
%for the global OCP does not hold for most real-word applications. 
see e.g.\ \cite{bib:zeilinger2013,bib:riverso2014,bib:riverso2015}, where
the OCP structure and size may change dynamically due to the removal
or plug-in of agents. 
The generation of the OCPs \eqref{eq:ADMM_step1} during run-time and the modular structure
of the framework allows to integrate
plug-and-play functionality in the network.
%has the further convenient
%consequence of enabling the concept of plug-and-play simulations
%\red{warum nur simulations?}, 
%For a static implementation this would have the
%consequence of changing the overall OCP while in the DMPC-framework
%changes can be applied dynamically. 
If an agent enters the system in the distributed control setting (left
part of Figure \ref{img:CommunicationStructure}), the coordinator
informs the direct neighbors of the agents to include the
corresponding variables. Hence, only the OCPs \eqref{eq:ADMM_step1} of the direct
neighbors are updated and the new agent is integrated into the network.
%while the OCPs of the remaining agents stay unchanged. 
If an agent leaves the network, the agents and the coordinator
delete the agent from their internal lists of active agents.
% and the simulation goes on. 
This results in a smooth transition between two problem formulations.

\section{Simulation examples}
\label{sec:simulation_example}

The modular framework \GRAMPCD is evaluated for different examples.
%In this section, simulation examples are presented that evaluate core
%characteristics of the framework. 
The scalability is shown for a coupled spring-mass system. 
%A \blue{scalable} system 
%An up-scaled system is simulated at
%first as a classical problem of coupled distributed systems. 
The plug-and-play functionality and the concept of neighbor
approximation are demonstrated for a smart grid and a coupled water
tank network, respectively.
%The second example demonstrates the plug-and-play functionality for a
%smart grid problem, before the concept of neighbor approximation is
%shown using coupled water tanks.
Finally, a distributed hardware implementation for a system of coupled Van der Pol oscillators is considered 
by communicating over an actual network using the TCP protocol. 

In each example, the terminal and integral cost in \eqref{eq:global_cost} are chosen quadratically
\begin{equation}
\begin{aligned}
V_i(\vec x_i, T) &=
\frac{1}{2}\norm{\vec x_i(T) -\vec x_{i,\text{des}}(T)}_{\vec P_i}^2 
\\
l_i(\vec x_i, \vec u_i, t) &= \frac{1}{2}\norm{\vec x_i - \vec
  x_{i,\text{des}}}_{\vec Q_i}^2 + \frac{1}{2}\norm{\vec u_i}_{\vec
  R_i}^2
\end{aligned}
\end{equation}
%
%$V_i(\vec x_i(T), T) =
%\frac{1}{2}\norm{\vec x_i(T)}_{\vec P_i}^2$, cost function $l_i(\vec
%x_i, \vec u_i, t) = \frac{1}{2}\norm{\vec x_i - \vec
%  x_{i,\text{des}}}_{\vec Q_i}^2 + \frac{1}{2}\norm{\vec u_i}_{\vec
%  R_i}^2$ 
with the positive (semi-)definite weighting
matrices $\vec P_i\in\R^{n_{x, i}\times n_{x, i}}$, $\vec Q_i
\in\R^{n_{x, i}\times n_{x, i}}$ and $\vec R_i \in\R^{n_{u, i}\times
  n_{u, i}}$. The desired state to be controlled is given by $\vec
x_{i,\text{des}}$. The computation times are measured on an Intel i5
CPU with $\SI{3.4}{\giga\hertz}$ using Windows 10. The communication effort is neglected if
not stated otherwise. 

\subsection{Scalable system}
\label{subsec:up-scaled-system}

\begin{figure}
	\centering
	\begin{tikzpicture}
		\begin{groupplot}
		[
			no markers,
			height = \heightPlot,
			width = \widthPlot
		]
		\nextgroupplot
		[
			xlabel=Simulation time  {$t\ [\si{\second}]$},
			ylabel=Cost \textit{J},
			ymax = 6500,
			ymin = 0,
			xmin = 0,
			xmax = 4
		]
		\addplot[line width = \linewidthfigure, blue] table [x=t, y=MPC, col sep=comma] {data/ssms/ssms.csv};
		\addplot[line width = \linewidthfigure, dashed, red ] table [x=t, y=DMPC, col sep=comma] {data/ssms/ssms.csv};
		\addlegendentry{ MPC };
		\addlegendentry{ DMPC };
		\end{groupplot}
	\end{tikzpicture}
	\caption{Global cost trajectory of the scalable spring-mass
          system with $40\times40$ agents for the central MPC and
          DMPC case.} 
	\label{fig:SSMS2D}
\end{figure}
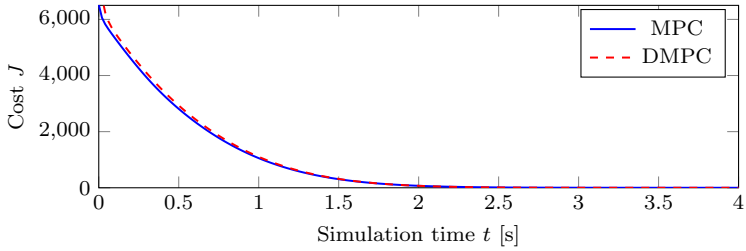

The scalability of \GRAMPCD is shown for a system consisting of a
set of masses that are coupled by springs. Each mass is represented by
an agent $i\in\V$ and is described by the differential equations 
\begin{align}
	\begin{bmatrix} \ddot p_{x, i} \\ \ddot p_{y, i} \end{bmatrix} =&\ \begin{bmatrix} u_{x, i} \\ u_{y, i} \end{bmatrix} + \sum_{j\in\N_i} \frac{c}{m} \left( 1 - \frac{\delta_0}{\delta_{ij}(p_{x, i}, p_{y, i})} \right) \begin{bmatrix} p_{x, j} - p_{x, i}  \\  p_{y, j} - p_{y, i}  \end{bmatrix}
	\label{eq:dynamics_ssms}%
\end{align}
with the position $(p_{x, i},p_{y, i})$ of the respective mass in the
$x-$ and $y$-axis and the respective controls $(u_{x, i}, u_{y,
  i})$. This results in the state and control vectors
\begin{subequations}
	\begin{align}
		\vec x_i =&\ \begin{bmatrix} p_{x, i} & \dot p_{x, i} & p_{y, i} & \dot p_{y, i} \end{bmatrix}^\TT \\
		\vec u_i =&\ \begin{bmatrix} u_{x, i} & u_{y, i} \end{bmatrix}^\TT.
	\end{align}
\end{subequations} 
The spring is relaxed at the length $\delta_0 =
\SI{1}{\meter}$. The spring constant is given by
$c=\SI{0.5}{\newton\per\meter}$ and each agent has a mass of $m_i =
\SI{7.5}{\kilogram}$. 
The function $\delta_{ij}(p_{x,i}, p_{y,i}) = \sqrt{ ( p_{x,i} - p_{x,j} )^2 + ( p_{y, i} -
  p_{y, j} )^2 }$ computes the distance between two agents $i$
and $j$. The dynamics \eqref{eq:dynamics_ssms} can be
split into functions $\vec f_i$ and $\vec f_{ij}$ corresponding to the 
neighbor-affine form \eqref{eq:probl:basic_OCP_dynamics}. The weighting matrices are set to 
(SI units are omitted for simplicity)
\begin{align}
\vec P_i = \diag{1, 1, 1, 1}\,,\quad \vec Q_i = \diag{5, 2, 5, 2}\,,\quad
\vec R_i = \diag{0.01, 0.01}
  \label{eq:ssms_weighting}%
\end{align}
with the desired state
\begin{align}
	\vec x_{i,\text{des}} = \begin{bmatrix}p_{x, i, \text{des}} & \SI{0}{\meter\per\second} & p_{y, i, \text{des}} & \SI{0}{\meter\per\second} \end{bmatrix}^\TT.
\end{align}
%
%The desired positions $p_{x, i, \text{des}}$ and $p_{y, i,
%  \text{des}}$ are ordered in a regular matrix form. The SI-units are
%omitted in \eqref{eq:ssms_weighting} for the sake of readability. 
%This system can be scaled by increasing the number of columns and rows of
%the matrix. 
It was shown in \cite{bib:burk2019} that the computation time 
per agent is nearly independent of the system size, whereas in the
central MPC case the computation time rises drastically. 
%It can be shown in numerical evaluations that for this
%system the computation time per agent is nearly independent of the
%system size in case of a distributed controller while for a
%centralized controller the computation time rises drastically, see
%e.g. \cite{bib:burk2019}. 
%A scalable system like this can be implemented into the DMPC framework
%very efficiently as the problem description can be described using
%loops. 
%Afterwards, both a 
%centralized and distributed controller can be used to solve the
%up-scaled system. 
The simulation results for a system with
$40\times40$ agents are given in Figure \ref{fig:SSMS2D}. It can be
seen that the trajectories of the cost are quite similar. 
While the distributed solution is slightly suboptimal, it would converge to the centralized solution by
increasing the number of \ADMM iterations.
The computation time for each time step, however, is
$\SI{1072.58}{\milli\second}$ for the centralized controller, while the
distributed controller requires a maximum of
$\SI{9.59}{\milli\second}$ and an average of
$\SI{2.23}{\milli\second}$ per agent. 
%This shows that distributed
%algorithms out-perform centralized approaches for up-scaled systems. 

\subsection{Plug-and-play}
\label{subsec:plug-and-play}
\newcommand*{\eleminvalid}{\ensuremath{\text{\scriptsize{\ding{54}}}}}

The plug-and-play capability of \GRAMPCD is presented using an
exemplary setup of a smart grid. The network is described by a set of
coupled agents that represent non-controllable power sinks and sources, such as
private households, industry or renewable energy as well as
controllable power plants. The dynamical behavior of the agents is
generalized by describing them as generators with a mechanical phase angle
$\theta_i(t)\in\R$ that may differ from the phase of the
grid. The corresponding dynamics 
\begin{align}
	\ddot \phi_i =&\ \frac{1}{I\Omega} \left( u_i + P_{\text{source}, i} - \kappa\Omega^2 \right) - 2\frac{\kappa}{I}\dot\phi - \sum_{j\in\N_i^\sending} \frac{P_{\text{max}, ij}}{I\Omega} \sin\left( \phi_j - \phi_i \right)
	\label{eq:pnp_dynamics}%
\end{align}
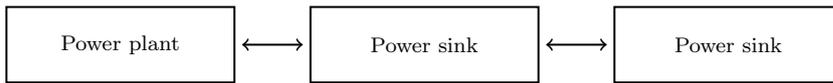
\begin{figure}
	\centering
	\begin{tikzpicture}
		\draw[line width = \linewidthfigure] (0, 0) rectangle (3, 1);
		\node at (1.5, 0.5) [] {Power plant};
		\draw[line width = \linewidthfigure, <->] (3.1, 0.5) -- (3.9, 0.5);
		\draw[line width = \linewidthfigure] (4, 0) rectangle (7, 1);
		\node at (5.5, 0.5) [] {Power sink};
		\draw[line width = \linewidthfigure, <->] (7.1, 0.5) -- (7.9, 0.5);
		\draw[line width = \linewidthfigure] (8, 0) rectangle (11, 1);
		\node at (9.5, 0.5) [] {Power sink};
	\end{tikzpicture}
	\caption{The plug-and-play capability of \GRAMPCD is
          presented using the simulation of a smart grid. At the
          beginning of the simulation, only one power sink is
          connected to the power plant. During run-time, the second
          power sink is connected to the first one, i.e.\ the power
          plant has to supply both using the same connection.} 
	\label{fig:smart_grid}
\end{figure}
with friction constant $\kappa>0$ and the moment of inertia $I$
is given in a neighbor-affine form and describes the dynamical
behavior of the phase shift $\phi_i(t)\in\R$ given by 
\begin{align}
	\phi_i = \theta_i - \Omega \tau
	\label{eq:pnp_theta}%
\end{align} 
between the phase $\Omega \tau$ of the grid with frequency $\Omega$ and the mechanical angle
$\theta_i$, see \cite{bib:rohden2012}. Hence, the state vector is given by
\begin{align}
	\vec x_i =&\ \begin{bmatrix} \phi_i & \dot \phi_i \end{bmatrix}^\TT .
\end{align}
In 
\eqref{eq:pnp_dynamics}, $P_{\text{source}, i}$ describes the generalized
non-controllable power, e.g. the demanded power for private households
and industry or the generated power by renewable energy. The
coupling between two agents consists of the maximum transferable power
$P_{\text{max}, ij}$ that depends on the phase shift angle
$\phi_j-\phi_i$ between agent $i\in\V$ and its neighbors
$j\in\N_i^\sending$. Agents that describe power plants have a
controllable input $u_i$ that is used to stabilize the grid. A normalized parameterization is used with 
$\Omega = \SI{1}{\hertz}$ and 
$I = \SI{1}{\joule\second\squared}$, the friction term set to
$\kappa=\SI{1e-3}{\joule\second}$ and the maximum transferable power to
$P_{\text{max}, ij}=\SI{0.1}{\joule\per\second}$. The weighting matrices are set to 
\begin{align}
	\vec P_i =&\ \diag{\SI{0}{\second\squared}, \SI{0.1}{\second\tothe{4}}}\,,& 
	\vec Q_i =&\ \diag{\SI{0}{\second\squared}, \SI{1}{\second\tothe{4}}}\,,& 
	\vec R_i =&\ \SI{0.01}{\second\squared\per\joule\squared}\,
\end{align}
with the desired state
\begin{align}
	\vec x_{i,\text{des}} = \begin{bmatrix} \times & \SI{0}{\per\second\squared} \end{bmatrix}^\TT.
\end{align}
The first element of the desired state $\vec x_{i, \text{des}}$
is set arbitrary, as 
%The first entry of the desired state is signed as \textit{dont-care}
%$\eleminvalid$ as 
there is no desired value for the
phase and the first state is not weighted in the cost
functional. 

\begin{figure}
	\centering
	\begin{tikzpicture}
		\begin{groupplot}
			[
				no markers,
				height = \heightPlot,
				width = \widthPlot,
				group style = {group size = 1 by 3},
				xmin = 0,
				xmax = 40
			]
			\nextgroupplot
			[
				ylabel=Global cost \textit{J},
				ymin = 0
			]
			\addplot[line width = \linewidthfigure, blue] table [x=t, y=cost, col sep=comma] {data/smart_grid/Cost.csv};
			\draw[line width = \linewidthfigure, <-] (axis cs: 22, 0.006) -- (25, 0.006);
			\node at (axis cs: 25, 0.006) [anchor=west, text width = 2cm]{Additional power sink is plugged in};
			\nextgroupplot
			[
				legend pos = south east,
				ylabel style={align=center},
				ylabel = Frequency \\ {$\dot\phi\ [\si{\radian\per\second}]$},
				ymax = 0
			]
			\addplot[ line width = \linewidthfigure, blue] table [x=t, y=f0, col sep=comma] {data/smart_grid/Frequenz0.csv};
			\addplot[ line width = \linewidthfigure, red, dashed ] table [x=t, y=f1, col sep=comma] {data/smart_grid/Frequenz1.csv};
			\addplot[ line width = \linewidthfigure, dotted ] table [x=t, y=f2, col sep=comma] {data/smart_grid/Frequenz2.csv};
			\addlegendentry{ Powerplant };
			\addlegendentry{ Power sink 1 };
			\addlegendentry{ Power sink 2 };
			\nextgroupplot
			[
				xlabel=Simulation time  {$t\ [\si{\second}]$},
				ylabel style={align=center, text width = 3cm},
				ylabel = {Phase shift \\ $\phi_i-\phi_j\ [\si{\radian}]$},
				legend pos = south west,
				ymax = 0
			]
			\addplot[ line width = \linewidthfigure, blue] table [x=t, y=diff01, col sep=comma] {data/smart_grid/Diff_01.csv};
			\addplot[line width = \linewidthfigure, red, dashed] table [x=t, y=diff12, col sep=comma] {data/smart_grid/Diff_12.csv};
			\addlegendentry{ Powerplant - Power sink 1 };
			\addlegendentry{ Power sink 1 - Power sink 2 };
		\end{groupplot}
	\end{tikzpicture}
	\caption{The Plug-and-play functionality of \GRAMPCD for the smart grid example is shown. The plot at the top shows the trajectory of the global cost, the plot in the middle the frequencies of the single agents and the plot in the bottom the phase shift between the agents. The additional power sink is plugged in at simulation time $t=\SI{20}{\second}$.}
	\label{fig:plot_smartGrid}
\end{figure}
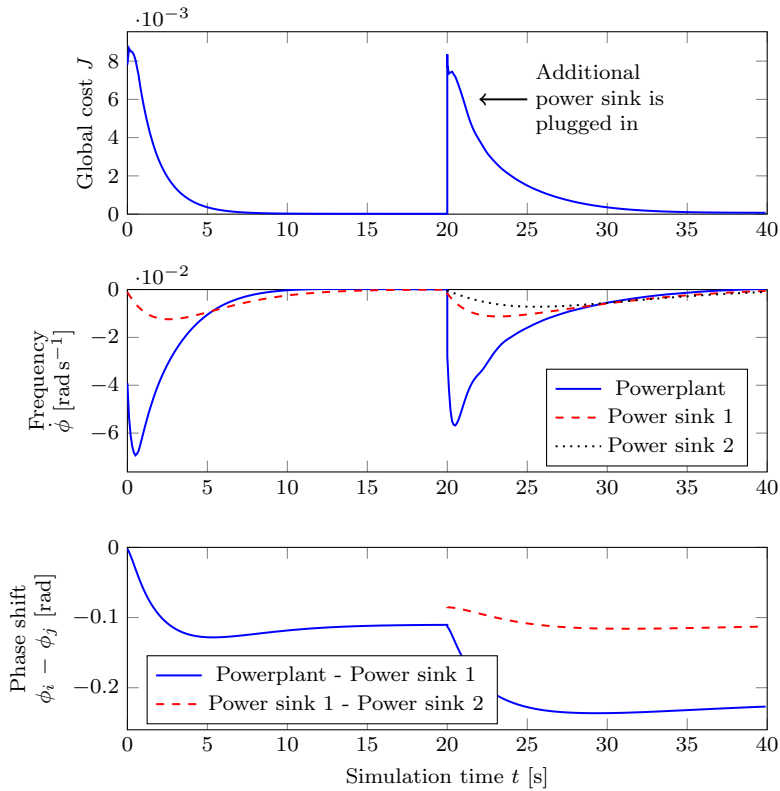

The implemented setup is visualized in Figure
\ref{fig:smart_grid}. At the start of the simulation, one power plant
is given that supplies one non-controllable power-sink such as a
private household. During run-time, an additional power sink is
coupled to the first one, i.e.\ the power
plant has to supply both using the same connection. 
The simulation results are
shown in Figure \ref{fig:plot_smartGrid}, starting with the first
power sink that is connected to the power
plant. It can be seen that the phase difference between the power
plant and the household converges to a stationary value that leads to
a transmission of the demanded power. Furthermore, the angular
velocity of the phase shift converges to zero. %$\SI{0}{\per\second}$, 
%hence the global cost converges as well. 
At simulation time $t=\SI{20}{\second}$, the second
power sink is plugged in, leading to an additional power demand. 
Consequently, the phase shift between the power plant and the first power sink
increases and the additional power is transmitted. The phase shift
between the first and second power sink is adapted accordingly. 
The computation time for the distributed controller is
given by a maximum of $\SI{84.67}{\milli\second}$ and an average of
$\SI{4.78}{\milli\second}$ per agent using a step size of $\Delta_t =
\SI{100}{\milli\second}$. The average time is significantly lower
due to the convergence criterion of the \ADMM\ algorithm.
% before
%reaching the maximum number of iterations if sufficient accuracy is
%reached. 

This plug-in and plug-out functionality of agents is supported at
any moment during the simulation even if the controllers run on
distributed hardware and communicate over a network. \GRAMPCD
can handle planned changes in the system such as shown in this
simulation example as well as spontaneous disconnections due to a broken
network connection. 
%This core functionality enables controlling
%systems with varying number of agents and couplings. 

\subsection{Neighbor approximation}
\label{subsec:neighborapproximation}

\begin{figure}[b]
	\centering
	\newcommand{\water}{blue!50!white}
	\begin{tikzpicture}
		% input
		\draw[line width = 0,fill = \water] (-0.5, 0.5) rectangle (0.25, 0.25);
		\draw[line width = \linewidthfigure] (-0.5, 0.5) -- (0.25, 0.5);
		\draw[line width = \linewidthfigure] (-0.5, 0.25) -- (0.25, 0.25);
		\draw[line width = \linewidthfigure, ->] (0.25, 0.375) -- (0.5, 0.375) -- (0.5, 0);
		% tank 1
		\draw[line width = \linewidthfigure, \water, fill = \water] (0, -0.25) rectangle (1, -1.5);
		\draw[line width = \linewidthfigure] (0, 0) -- (0, -1.5) -- (1, -1.5) -- (1, 0);
		\node at (0.5, -0.75) {$1$};
		\draw (0.5, -0.75) circle (0.375cm);
		% pipe 1-2
		\draw[line width = \linewidthfigure, fill=\water] (1, -1.25) rectangle (2, -1);
		% tank 2
		\draw[line width = \linewidthfigure, \water, fill = \water] (2, -0.375) rectangle (3, -1.5);
		\draw[line width = \linewidthfigure] (2, 0) -- (2, -1.5) -- (3, -1.5) -- (3, 0);
		\node at (2.5, -0.75) {$2$};
		\draw (2.5, -0.75) circle (0.375cm);
		% pipe 2-3
		\draw[line width = \linewidthfigure, fill=\water] (3, -1.25) rectangle (4, -1);
		% tank 3
		\draw[line width = \linewidthfigure, \water, fill = \water] (4, -0.5) rectangle (5, -1.5);
		\draw[line width = \linewidthfigure, ] (4, 0) -- (4, -1.5) -- (5, -1.5) -- (5, 0);
		\node at (4.5, -0.75) {$3$};
		\draw (4.5, -0.75) circle (0.375cm);
		% pipe 3-4
		\draw[line width = \linewidthfigure, fill=\water] (5, -1.25) rectangle (6, -1);
		% tank 4
		\draw[line width = \linewidthfigure, \water, fill = \water] (6, -0.625) rectangle (7, -1.5);
		\draw[line width = \linewidthfigure] (6, 0) -- (6, -1.5) -- (7, -1.5) -- (7, 0);
		\node at (6.5, -0.75) {$4$};
		\draw (6.5, -0.75) circle (0.375cm);
		% pipe 4-5
		\draw[line width = \linewidthfigure, fill=\water] (7, -1.25) rectangle (8, -1);
		% tank 5
		\draw[line width = \linewidthfigure, \water, fill = \water] (8, -0.75) rectangle (9, -1.5);
		\draw[line width = \linewidthfigure] (8, 0) -- (8, -1.5) -- (9, -1.5) -- (9, 0);
		\node at (8.5, -0.75) {$5$};
		\draw (8.5, -0.75) circle (0.375cm);
		% pipe 5-
		\draw[line width = 0, fill=\water, fill = \water] (9, -1) rectangle (9.5, -1.25);
		\draw[line width = \linewidthfigure, fill=\water] (9, -1) -- (9.5, -1);
		\draw[line width = \linewidthfigure, fill=\water] (9, -1.25) -- (9.5, -1.25);
		\draw[line width = \linewidthfigure, ->] (9.5, -1.125) -- (10, -1.125) -- (10, -1.5);
	\end{tikzpicture}
	\caption{The concept of neighbor approximation is shown at a simulation example of coupled water tanks. Only the first one has an input and only the last one has a desired water height while being disturbed by a constant outflow.}
	\label{img:img_watertank}
\end{figure}
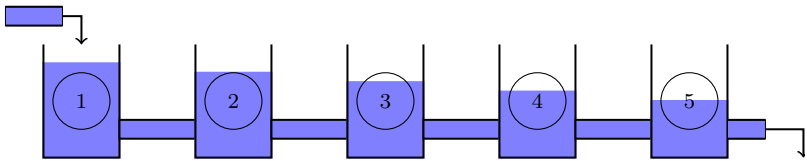

\begin{figure}
	\centering
	\begin{tikzpicture}
		\begin{groupplot}
			[
				no markers,
				height = 1.2*\heightPlot,
				width = \widthPlot
			]
			\nextgroupplot
			[
				xlabel = \ADMM\ iterations,
				ylabel = {Cost $J$},
				legend pos = south east,
				ymin = 8,
				xmin = 1,
				xmax = 89
			]
			\addplot[line width = \linewidthfigure, dashed, red] table [x = iter, y=cost, col sep=comma] {data/waterTank/cost_with.csv};
			\addplot[line width = \linewidthfigure, blue] table [x=iter, y=cost, col sep=comma] {data/waterTank/cost_without.csv};
			\addlegendentry{With neighbor approximation}
			\addlegendentry{Without neighbor approximation}
		\end{groupplot}
	\end{tikzpicture}
	\caption{Convergence behavior of the \ADMM\ algorithm with and without neighbor approximation. The algorithm converges within $89$ \ADMM\ iterations without neighbor approximation opposed to $7$ iterations if neighbor approximation is used.}
	\label{fig:plot_waterTank_without}
\end{figure}
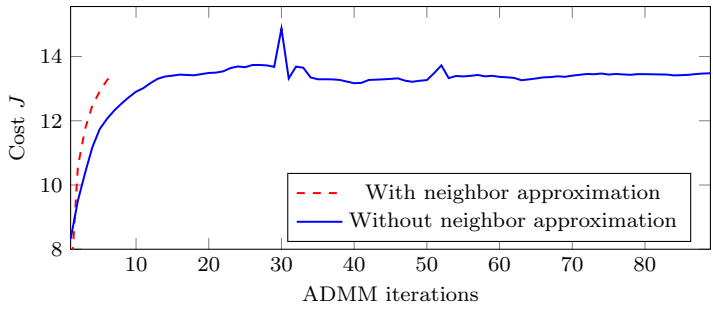

The concept of neighbor approximation is evaluated for a system of water tanks that are coupled by pipes, see Figure
\ref{img:img_watertank}. Only the first water tank has a controllable
input $u_1(t)\in\R$. The last water tank has a constant outflow $d_5 =
\SI{0.01}{\meter\cubed\per\second}$ and the
desired water height $x_{5, des}=\SI{3}{\meter}$.
% as well as a
%distortion in form of a constant outflow $d_5 =
%\SI{0.01}{\meter\cubed\per\second}$. 
In addition, the inequality constraint $h_i\leq
\SI{3}{\meter}$ of a maximum water height has to be satisfied by all
tanks. The dynamics of each water tank is given by 
\begin{align}
	\dot h_i =&\ \frac{1}{A_i} \left( u_i - d_i \right) + \sum_{j\in\N_i^\sending} \frac{a_{ij}}{A_i} \sign{h_j-h_i} \sqrt{2g\abs{h_j-h_i}}
\end{align}
with the water height $h_i(t)\in\R$ and the state vector $x_i = h_i$.
%and the difference $\delta_{h,
%  ij} = h_j - h_i$ of the water height of two agents $i$ and $j$. 
The area of each water tank is given by $A_i= \SI{0.1}{\meter\squared}$
and the diameter of the pipes by
$a_{ij}=\SI{0.005}{\meter\squared}$. The weights for the cost
functions are set to 
\begin{subequations}
	\begin{align}
		P_1 =&\ \SI{0}{\per\meter\squared},& 
		Q_1 =&\ \SI{0}{\per\meter\squared},& 
		R_1 =&\ \SI{0.1}{\second\squared\per\meter\tothe{6}}\\
		P_i =&\ \SI{0}{\per\meter\squared},& 
		Q_i =&\ \SI{0}{\per\meter\squared},& 
		R_i =&\ \SI{0}{\second\squared\per\meter\tothe{6}},\quad 
		i\in\{2, 3, 4\}\\
		P_5 =&\ \SI{1}{\per\meter\squared},& 
		Q_5 =&\ \SI{1}{\per\meter\squared},& 
		R_5 =&\ \SI{0}{\second\squared\per\meter\tothe{6}}.
	\end{align}
\end{subequations} 
The simulation is run with a distributed controller both with and
without neighbor approximation using the same set of parameters for
the \ADMM\ algorithm and \GRAMPC. The convergence of the \ADMM\
algorithm is shown in Figure \ref{fig:plot_waterTank_without} for both
simulations. 
%The convergence of the standard \ADMM\ algorithm is given
%with a dashed, blue line. 
%It can be seen, that the algorithm does not
%converge within the first $100$ iterations. 
The simulation with neighbor approximation 
%is presented in a solid, red line. In this case the algorithm 
converges smoothly to the
optimal solution and satisfies the convergence criterion after $7$
iterations while $89$ \ADMM iterations are required without neighbor approximation. Note that the cost is rising instead of falling as the solution is infeasible
until the algorithm converges. The improved convergence behavior with neighbor approximation comes with a higher computational
complexity per \ADMM\ iteration. However, if a convergence criterion
is used, the decreased number of \ADMM\ iterations per time step
compensate for the higher computational complexity. The required
computation time for the $89$ \ADMM\ iterations without neighbor
approximation is $\SI{955.6}{\milli\second}$ and for the $7$
iterations using neighbor approximation
$\SI{203.8}{\milli\second}$. 
%Hence, a better solution results with less computation effort. 
Note that the same configuration for the
\ADMM\ algorithm and \GRAMPC\ is used in this evaluation to provide a
comparable result. The standard \ADMM\ algorithm may converge within
less iterations if more computation effort is spent per iteration
while the algorithm may require even less time if the parameters are
tuned for neighbor approximation. 

\subsection{Distributed hardware implementation}
\label{subsec:distributedOptimization}

This section shows the capability of \GRAMPCD to solve the
\ADMM\ algorithm on distributed embedded hardware. 
The following simulation is run on four Raspberry Pi $3\mathrm B +$ using Raspberry Pi OS
that are connected via Ethernet. Each agent runs on an
individual Raspberry Pi while the coordinator and the simulator use
the same hardware. The local communication interface from \GRAMPCD based on the TCP protocol is used. 
The simulation example consists of three
coupled Van der Pol oscillators~\cite{bib:barron2009}
\begin{align}
	\ddot p_i =&\ \alpha_1 \left( 1 - p_i^2 \right) - p_i + u_i + \sum_{j\in\N_i^\sending} \alpha_2 \left( p_j - p_i \right)
\end{align}
with state $p_i(t)\in\R$, control $u_i(t)\in\R$,
the uncoupled
oscillator constant $\alpha_1=\SI{1}{\per\meter\per\second\squared}$,
and the coupling constant $\alpha_2=\SI{1}{\per\second\squared}$. 
%Each agent can
%additionally accelerate its displacement using the controllable input
%$u_i(\tau)\in\R$. 
The state vector and desired state are given by
\begin{subequations}
	\begin{align}
		\vec x_i =&\ \begin{bmatrix} p_i & \dot p_i \end{bmatrix}^\TT \\
		\vec x_{i, \text{des}} =&\ \begin{bmatrix} \SI{0}{\meter} & \SI{0}{\meter\per\second} \end{bmatrix}^\TT
	\end{align}
\end{subequations}
with the weighting matrices set to
\begin{subequations}
	\begin{align}
		\vec P_i =&\ \mathrm{diag} \begin{bmatrix} \SI{1}{\per\meter\squared} & \SI{1}{\per\meter\squared\second\squared} \end{bmatrix},\\
		\vec Q_i =&\ \mathrm{diag} \begin{bmatrix} \SI{1}{\per\meter\squared} & \SI{1}{\per\meter\squared\second\squared} \end{bmatrix},\\
		R_i =&\ \SI{0.1}{\per\meter\squared\second\tothe{4}}.
	\end{align}
\end{subequations}
The simulation is run for $10.000$ time steps using
a fixed number of $q_\text{max}= 5$ \ADMM\ iterations. 

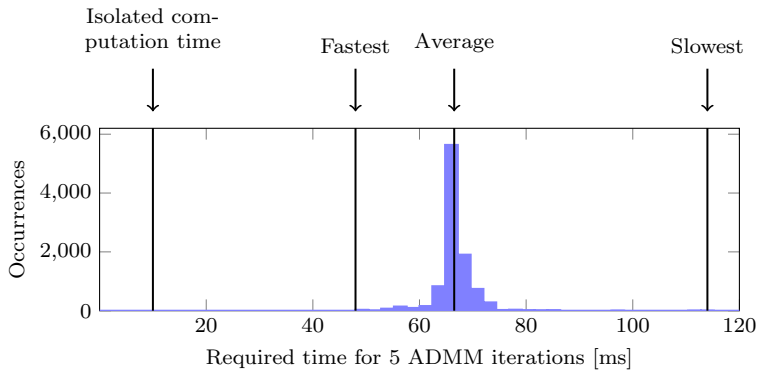
\begin{figure}[h]
	\centering
	\newcommand{\maximum}{6500}
	\newcommand{\minvalue}{10}
	\newcommand{\fastest}{48}
	\newcommand{\avgvalue}{66.51}
	\newcommand{\slowest}{114}
	\newcommand{\lightblue}{blue!50!white}
	\newcommand{\topvalue}{8500}
	\begin{tikzpicture}
		\begin{axis}
		[
			no markers,
			xtick = {20, 40, 60, 80, 100, 120},
			xmin = 0,
			xmax = 120,
			ymin = 0,
			height = \heightPlot,
			width = \widthPlot,
			xlabel = {Required time for $5$ \ADMM\ iterations $[\si{\milli\second}]$},
			ylabel = Occurrences
		]
		\addplot+[line width = \linewidthfigure, hist={bins=50}, \lightblue, fill = \lightblue] table [y index=0]{data/distributed/Times.dat};
		\draw[line width = \linewidthfigure] (axis cs: \minvalue, 0) -- (axis cs: \minvalue, \maximum);
		\draw[line width = \linewidthfigure] (axis cs: \fastest, 0) -- (axis cs: \fastest, \maximum);
		\draw[line width = \linewidthfigure] (axis cs: \avgvalue, 0) -- (axis cs: \avgvalue, \maximum);
		\draw[line width = \linewidthfigure] (axis cs: \slowest, 0) -- (axis cs: \slowest, \maximum);
		\node (comp) at (axis cs: \minvalue, \maximum) {};
		\node (compTop) at (axis cs: \minvalue, \topvalue) {};
		\node (fastest) at (axis cs: \fastest, \maximum) {};
		\node (fastestTop) at (axis cs: \fastest, \topvalue) {};
		\node (avg) at (axis cs: \avgvalue, \maximum) {};
		\node (avgTop) at (axis cs: \avgvalue, \topvalue) {};
		\node (slowest) at (axis cs: \slowest, \maximum) {};
		\node (slowestTop) at (axis cs: \slowest, \topvalue) {};
		\end{axis}
		\draw[line width = \linewidthfigure, ->] (compTop) -- (comp);
		\node at (compTop) [anchor=south, text width = 2cm, align=center] {Isolated computation time};
		\draw[line width = \linewidthfigure, ->] (fastestTop) -- (fastest);
		\node at (fastestTop) [anchor=south, text width = 3cm, align=center] {Fastest};
		\draw[line width = \linewidthfigure, ->] (avgTop) -- (avg);
		\node at (avgTop) [anchor=south, text width = 2cm, align=center] {Average};
		\draw[line width = \linewidthfigure, ->] (slowestTop) -- (slowest);
		\node at (slowestTop) [anchor=south, text width = 3cm, align=center] {Slowest};
	\end{tikzpicture}
	\caption{Communication effort of the \ADMM\ algorithm on
          distributed hardware with 
          $\SI{48}{\milli\second}$ (minimum), 
          $\SI{66.51}{\milli\second}$ (average), and 
          $\SI{114}{\milli\second}$ (maximum) compared to the
          computation time of $\SI{10}{\milli\second}$. 
          %This shows that the main effort
          %in \DMPC\ is the communication as the required time for the
          %computation is given by $\SI{10}{\milli\second}$.
        } 
	\label{img:distributed_plot}
\end{figure}

Figure \ref{img:distributed_plot} shows the required time for
computation and communication in each time step. The 
computation time to execute $5$ \ADMM\ iterations amounts to
$\SI{10}{\milli\second}$ per agent while the average time required to
solve the \ADMM\ algorithm in a distributed manner using the TCP
protocol for the communication is $\SI{66.51}{\milli\second}$. Hence,
the average effort for the communication is
$\SI{56.51}{\milli\second}$ including $72$ communication steps. 
Since all agents have to be synchronized for the \ADMM\ algorithm,
either of them has to wait for the slowest agent at each step of the
algorithm. All five \ADMM\ iterations can be executed in
$\SI{48}{\milli\second}$ with the worst case time of 
$\SI{114}{\milli\second}$. This results in the minimum time for each
communication step of $\SI{0.53}{\milli\second}$, an average time of
$\SI{0.79}{\milli\second}$ and a maximum time of
$\SI{1.44}{\milli\second}$. This time includes preparing the data to
be sent, sending and receiving it and recreating the sent data
structure from the byte array. These are plausible values as a ping to
the loopback address $127.0.0.1$ already requires
$\SI{0.14}{\milli\second}$ in average and $\SI{0.22}{\milli\second}$
at maximum. These results show that the main effort in distributed
optimization is the communication effort that requires $82.3\%$ of the
overall time in average for this simulation example.

\section{Conclusions}
\label{sec:conclusions}

The open-source, modular DMPC framework \GRAMPCD is presented in this paper
that enables to solve scalable
optimal control problems in a convenient way and to stabilize plants using distributed model predictive control. This problem
description can be used for both a centralized and a distributed
controller. 
%In the first case, the global
%optimal control problem is 
%generated and solved using the MPC-toolbox \GRAMPC, while in the
%second 
In the distributed setting, the global optimal control problem is
automatically decoupled and solved in a distributed manner using the \ADMM\
algorithm. The convergence behavior of the \ADMM\
algorithm can be improved by sugint he concept of neighbor approximation that allows the agents to anticipate the actions of their neighbors. The presented DMPC framework supports plug-and-play to
connect and remove agents at run-time. 
Besides solving the \ADMM\ algorithm on a single processor, it is
possible to solve the local optimal control problems on distributed
hardware. The local communication interface enables communication
between agents over a network using the TCP protocol. 
By default, \GRAMPCD uses the MPC toolbox \GRAMPC\ for
solving the local optimal control problems on agent level, which is
suitable for real-time and embedded implementations.

\GRAMPCD is licensed under the Berkeley Software Distribution 3-clause version (BSD-3) license. The complete source-code is available at Github \linebreak\href{https://github.com/grampc-d/grampc-d}{https://github.com/grampc-d/grampc-d}.
Future work will use the modular structure to extend \GRAMPCD. For example, communication protocols besides TCP can
be provided or alternative solvers to \GRAMPC\ for the underlying
minimization problem implemented to increase the usability and
flexibility of the framework.

\begin{acknowledgements}
	This work was funded by the Deutsche Forschungsgemeinschaft (DFG, German Research Foundation) under project no.	GR 3870/4-1.
\end{acknowledgements}

\bibliographystyle{IEEEtran}
\bibliography{dmpc_literature} 

\end{document}